\documentclass[11pt]{article}
\usepackage{amsmath,amsfonts,amsthm,amssymb,ifthen,epsfig,subfigure,caption}

\newboolean{@laptop}
\newboolean{@todoon}
\setboolean{@laptop}{true}
\setboolean{@todoon}{true}

\ifthenelse{\boolean{@laptop}}{}{\usepackage{mathbbol}}

\def\todo#1{\ifthenelse{\boolean{@todoon}}{\marginpar{\textit{#1}}}{}}

\newtheorem{theorem}{Theorem}[section]

\newtheorem{lemma}[theorem]{Lemma}

\newtheorem{proposition}[theorem]{Proposition}

\newtheorem{claim}[theorem]{Claim}

\newcommand{\Furedi}{F{\"u}redi}
\newcommand{\Komlos}{Koml{\'o}s}

\newcommand{\sparsify}{\ensuremath{\mathtt{Sparsify}}}
\newcommand{\sparsifytwo}{\ensuremath{\mathtt{Sparsify2}}}
\newcommand{\unwtedsparsify}{\ensuremath{\mathtt{UnwtedSparsify}}}
\newcommand{\boundedsparsify}{\ensuremath{\mathtt{BoundedSparsify}}}

\newcommand{\sample}{\ensuremath{\mathtt{Sample}}}
\newcommand{\approxcut}{\ensuremath{\mathtt{ApproxCut}}}

\newcommand{\partition}{\ensuremath{\mathtt{Partition}}}
\newcommand{\partitiontwo}{\ensuremath{\mathtt{Partition2}}}
\newcommand{\partsample}{\ensuremath{\mathtt{PartitionAndSample}}}

{\begin{list}{}{\usecounter{bean}%
  \setlength{\topsep}{0em}
  \setlength{\parsep}{0em}
  \setlength{\partopsep}{0em}
  \setlength{\itemsep}{0em}
}}%
{\end{list}}

\def\calL{\mathcal{L}}

\def\Htil{\widetilde{H}}
\def\Itil{\widetilde{I}}
\def\Dtil{\widetilde{D}}
\def\dtil{\tilde{d}}

\def\Gtil{\widetilde{G}}
\def\Etil{\widetilde{E}}
\def\Ftil{\widetilde{F}}
\def\Ltil{\widetilde{L}}

\def\Atil{\widetilde{A}}

\def\Ghat{\widehat{G}}

\def\wtil{\tilde{w}}
\def\ytil{\tilde{y}}

\def\ztil{\tilde{z}}

\def\vs#1#2#3{#1_{#2},\ldots , #1_{#3}}

\def\phat{\hat{p}}
\def\epshat{\hat{\epsilon}}

\def\softO#1{\widetilde{\mathcal{O}} \left( #1 \right)}
\def\bigO#1{\widetilde{\mathcal{O}} \left( #1 \right)}

\def\myPhiSym{\ifthenelse{\boolean{@laptop}}%
{\varphi}
{\mathbb{\Phi}}
}

\def\trace#1{\mathrm{Tr} \left(#1 \right)}

\def\bdry#1#2{\partial_{#1}\left(#2\right)}

\def\pleq{\preccurlyeq}
\def\pgeq{\succcurlyeq}

\def\union{\cup}
\def\intersect{\cap}

\def\defeq{\stackrel{\mathrm{def}}{=}}

\def\prob#1#2{\Pr_{#1}\left[ #2 \right]}
\def\expec#1#2{\mbox{\bf E}_{#1}\left[ #2 \right]}

\def\norm#1{\left\| #1 \right\|}

\def\setof#1{\left\{#1  \right\}}
\def\sizeof#1{\left|#1  \right|}

\def\bvec#1{{\mbox{\boldmath $#1$}}}

\def\pleq{\preccurlyeq}
\def\pgeq{\succcurlyeq}

\def\abs#1{\left|#1  \right|}
\def\intersect{\cap}

\newcommand{\ceiling}[1]{\lceil#1\rceil}

\def\bigO#1{O\left(#1  \right)}

\def\setof#1{\left\{#1  \right\}}
\def\abs#1{\left|#1  \right|}
\def\vol#1{\mathrm{Vol}\left(#1  \right)}

\newdimen\pIR
\newcommand\StevesR{{\rm I\kern\pIR R}}
\def\Reals#1{\StevesR^{#1}}

\def\conduc#1#2{\Phi_{#1}\left(#2  \right)}
\def\conducin#1#2{\Phi^{G}_{#1}\left(#2  \right)}

\def\Conduc#1{\Phi_{#1}}
\def\Conducin#1{\Phi^{G}_{#1}}

\def\blowup#1#2{\textrm{blow-up}_{#1}\left( #2 \right)}

\def\preDelta#1{{}^{\delta}\!{#1}}
\begin{document}

\title{Spectral Sparsification of Graphs%
\thanks{
This paper is the second in a sequence of three papers expanding
  on material that appeared first under the title
  ``Nearly-linear time algorithms for graph partitioning, 
    graph sparsification, and solving linear systems''~\cite{SpielmanTengPrecon}.
The first paper,
``A Local Clustering Algorithm for Massive Graphs and its Application to Nearly-Linear Time Graph Partitioning''~\cite{SpielmanTengCuts}
 contains graph partitioning algorithms that are used to construct the sparsifiers in this paper.
The third paper, ``Nearly-Linear Time Algorithms for Preconditioning and Solving Symmetric, Diagonally Dominant Linear Systems''~\cite{SpielmanTengLinsolve} contains the results
  on solving linear equations and approximating eigenvalues and eigenvectors.
\vskip 0.01in
This material is based upon work supported by the National Science Foundation 
  under Grant Nos. 0325630, 0324914, 0634957, 0635102 and 0707522.
Any opinions, findings, and conclusions or recommendations expressed in this material are those of the authors and do not necessarily reflect the views of the National Science Foundation.
\vskip 0.01in
Shang-Hua Teng wrote part of this paper while at MSR-NE lab and Boston
University.}
}

\author{
Daniel A. Spielman\\
Department of Computer Science\\
Program in Applied Mathematics\\
Yale University
\and
Shang-Hua Teng\\
Department of Computer Science\\
Viterbi School of Engineering\\
University of Southern California}

\maketitle

\begin{abstract}
We introduce a new notion of graph sparsification based on 
  spectral similarity of graph Laplacians:
 spectral sparsification requires that the Laplacian quadratic form
  of the sparsifier approximate that
  of the original.
This is equivalent to saying that the Laplacian of the sparsifier
  is a good preconditioner for the Laplacian of the original.

We prove that every graph has a spectral sparsifier of nearly-linear size.
Moreover, we present an algorithm that produces spectral sparsifiers
   in time $\bigO{m \log^{c} m}$, where $m$ is the number of edges in the
   original graph and $c$ is some absolute constant.
This construction is a key component of a nearly-linear time algorithm
  for solving linear equations in diagonally-dominant matrices.

Our sparsification algorithm makes use of a nearly-linear time algorithm for graph
  partitioning that satisfies a strong guarantee:
 if the partition it outputs is very unbalanced, then the larger part is contained in
  a subgraph of high conductance.
\end{abstract}

\newpage

\section{Introduction}
Graph sparsification is the task of approximating a graph by a sparse graph,
  and is often useful in the design of efficient approximation algorithms.
Several notions of graph sparsification have been proposed.
For example, Chew \cite{PaulChew}
  was motivated by  proximity problems in computational geometry
  to introduce graph spanners.
Spanners are defined in terms of the {\em distance similarity}
  of two graphs:
A spanner is a sparse graph in which the shortest-path distance 
  between every pair of vertices is approximately the same in the original graph
  as in the spanner.
Motivated by cut problems, Benczur and Karger~\cite{BenczurKarger} introduced
  a notion of sparsification that requires that for every set of vertices,
  the weight of the edges
  leaving that set should be approximately the same 
  in the original graph as in the sparsifier.

Motivated by problems in numerical linear algebra and spectral graph theory,
  we introduce a new notion of sparsification 
  that we call \textit{spectral sparsification}.
A spectral sparsifier is a subgraph of the original whose Laplacian
  quadratic form is approximately the same as that of the original graph
  on all real vector inputs.
The Laplacian matrix\footnote{For more information on the Laplacian matrix of a graph, we refer the
  reader to one of~\cite{BollobasMGT,Mohar91Laplacian,GodsilRoyle,Chung}.} of a weighted graph $G = (V,E,w)$, where $w_{(u,v)}$ is the weight of edge $(u,v)$, is defined by
\[
  L_{G} (u,v) =
\begin{cases}
-w_{(u,v)} & \text{if $u \not = v$}
\\
\sum_{z} w_{ (u,z)} & \text{if $u = v$}.
\end{cases}
\]
It is better understood by its quadratic form, which on $x \in \Reals{V}$
  takes the value
\begin{equation}\label{eqn:quadraticForm}
  x^{T} L_{G} x = \sum_{(u,v) \in E} w_{(u,v)} \left(x (u) - x (v) \right)^{2}.
\end{equation}

We say that $\Gtil$ is a { \em $\sigma$-spectral approximation}
   of $G$ if for all $x \in \Reals{V}$
\begin{equation}\label{eqn:approxForm}
  \frac{1}{\sigma} x^{T} L_{\Gtil } x \leq x^{T} L_{G} x \leq \sigma x^{T} L_{\Gtil } x.
\end{equation}

Our notion of sparsification captures the {\em spectral similarity}
  between a graph and its sparsifiers.
It is  a stronger notion than the
  cut sparsification of Benczur and Karger:
 the cut-sparsifiers constructed by Benczur and Karger~\cite{BenczurKarger}
  are only required to satisfy these inequalities for all $x \in \setof{0,1}^{V}$.
In Section~\ref{sec:examples} we present an example demonstrating that these notions
  of approximation are in fact different.

Our main result is
  that every weighted graph has a spectral sparsifier with
  $\softO{n}$ edges that can be computed in
  $\softO{m}$ time,
 where we recall that
  $\softO{f (n)}$
  means $O (f (n) \log^{c} f (n))$, for some constant $c$.
In particular, we prove that for every weighted graph $G = (V,E,w)$ and every
  $\epsilon > 0$, there is a re-weighted subgraph of $G$ with $\softO{n / \epsilon^{2}}$ edges
  that is a $(1+\epsilon)$ approximation of $G$.
Moreover, we show how to  find such
  a subgraph in $\softO{m}$ time, where $n = \sizeof{V}$ and $m = \sizeof{E}$.
The constants and powers of logarithms
  hidden in the $\widetilde{\mathcal{O}}$-notation
  in the statement of our results are quite large.
Our goal in this paper is not to produce sparsifiers with optimal parameters, but
  rather just to prove
  that spectral sparsifiers with a nearly-linear number of edges exist and
  that they can be found in nearly-linear time.

Our sparsification algorithm makes use of a nearly-linear time
  graph partitioning algorithm, \approxcut ,  that we develop in Section~\ref{sec:approxCut} and which may be of independent interest.
On input a target conductance $\phi$, \approxcut \ always outputs a set of vertices
  of conductance less than $\phi$.
With high probability, if the set it outputs is small then its complement
 is contained in a subgraph of conductance at least $\Omega (\phi^{2} / \log^{4} m)$.

\section{The Bigger Picture}

This paper arose in our efforts to design nearly-linear time algorithms
  for solving diagonally-dominant linear systems, and is the second in
  a sequence of three papers on the topic.
In the first paper~\cite{SpielmanTengCuts}, we develop fast routines
  for partitioning graphs, which we then use in our algorithms for building
  sparsifiers.
In the last paper~\cite{SpielmanTengLinsolve}, we show how to use
  sparsifiers to build preconditioners for diagonally-dominant matrices
  and thereby solve linear equations in such matrices in nearly-linear time.
Koutis, Miller and Peng~\cite{KMP} have recently developed an algorithm for
  solving such systems 
  of linear equations in time $O (m \log^{2} n)$
  that does not rely upon the sparsifiers of the present paper.

The quality of a preconditioner is measured by the relative condition number,
  which for the Laplacian matrices of a graph $G$ and its sparsifier $\Gtil$ is
\[
\kappa (G, \Gtil) \defeq
 \left( \max_{x} \frac{x^{T} L_{G} x}{x^{T} L_{\Gtil } x} \right)
\Big/
 \left(  \min_{x} \frac{x^{T} L_{G} x}{x^{T} L_{\Gtil } x} \right)
\]
So, if $\Gtil$ is a $\sigma $-spectral approximation of $G$ then
  $\kappa (G, \Gtil) \leq \sigma^{2}$.
This means that
  an iterative solver such as the Preconditioned Conjugate Gradient~\cite{Axelsson}
 can solve a linear system in the Laplacian of
  $G$ to accuracy $\epsilon$ by solving $O (\sigma \log (1/\epsilon ))$
  linear systems in
  $\Gtil$ and performing as many multiplications by $G$.
As a linear system in a matrix with $m$ non-zero entries may be solved
  in time $O (n m)$ by using the Conjugate Gradient as a
  direct method~\cite[Theorem~28.3]{TrefethenBau},
 the use of the sparsifiers in this paper alone provides an
  algorithm for solving linear systems in $L_{G}$ to $\epsilon$-accuracy in
  time $\softO{n^{2} \log (1/\epsilon)}$, which is nearly optimal when
  the Laplacian matrix has $\Omega(n^2)$ non-zero entries.
In our paper on solving linear equations~\cite{SpielmanTengLinsolve}, we show
  how to get the time bound down to $\softO{m \log (1/\epsilon)}$,
  where $m$ is the number of non-zero entries in $L_{G}$.


\section{Outline}
In Section~\ref{sec:background}, we present technical background required
  for this paper, and maybe even for the rest of this outline.
In Section~\ref{sec:examples}, we present three examples of graphs and
  their sparsifiers.
These examples help motivate key elements of our construction.

There are three components to our algorithm for sparsifying graphs.
The first is a random sampling procedure.
In Section~\ref{sec:sampling}, we prove that this procedure produces
  good spectral sparsifiers for graphs of high conductance.
So that we may reduce the problem of sparsifying arbitrary graphs
  to that of sparsifying graphs of high conductance, we require
  a fast algorithm for partitioning a graph into parts of high conductance
  without removing too many edges.
In Section~\ref{sec:decomp}, we first prove that such partitions exist,
  and use them to prove the existence of spectral sparsifiers for
  all unweighted graphs.
In Section~\ref{sec:approxCut}, we then build on tools from~\cite{SpielmanTengCuts}
  to develop a graph partitioning procedure that suffices.
We use this procedure in Section~\ref{sec:unweighted} to construct
  a nearly-linear time algorithm
  for sparsifying unweighted graphs.
We show how to use this algorithm to sparsify
  weighted graphs in Section~\ref{sec:weighted}.

We conclude in Section~\ref{sec:conclusion} by surveying recent improvements
  that have been made in both sparsification and in the
  partitioning routines on which the present paper depends.

\section{Background and Notation}\label{sec:background}
By $\log$ we always mean the logarithm base $2$, and we denote the
  natural logarithm by $\ln$.

As we spend this paper studying spectral approximations, we will say
  ``$\sigma$-approximation'' instead of 
  ``$\sigma$-spectral approximation'' wherever it won't create
  confusion.

We may express \eqref{eqn:approxForm} more compactly by employing the notation
  $A \pleq B$ to mean
\[
  x^{T} A x \leq x^{T} B x, \quad \text{for all $x \in \Reals{V}$}.
\]
Inequality $\eqref{eqn:approxForm}$ is then equivalent to
\begin{equation}\label{eqn:approxPleq}
  \frac{1}{\sigma} L_{\Gtil} \pleq L_{G} \pleq \sigma L_{\Gtil}.
\end{equation}
We will overload notation by writing $G \pleq \Gtil $ for graphs $G$ and $\Gtil$
  to mean $L_{G} \pleq L_{\Gtil}$.

For two graphs $G$ and $H$, we write
\[
  G + H
\]
to indicate the graph whose Laplacian is $L_{G} + L_{H}$.
That is, the weight of every edge in $G+H$ is the sum of the weights of the
  corresponding edges in $G$ and $H$.
We will use this notation even if $G$ and $H$ have different vertex sets.
For example,
  if their vertex sets are disjoint, then their sum is simply
  the disjoint union of the graphs.
It is immediate that $G \pleq \Gtil$ and $H \pleq \Htil$ imply
\[
  G + H \pleq \Gtil + \Htil.
\]
In many portions of this paper, we will consider vertex-induced subgraphs of graphs.
\textit{When we take subgraphs, we always preserve the identity of vertices.}
This enables us to sum inequalities on the different subgraphs to say something
  about the original.

For an unweighted graph $G = (V,E)$, we will let $d_{v}$ denote the degree
  of vertex $v$.
For $S$ and $T$ disjoint subsets of $V$, we let $E (S,T)$ denote the set of edges
  in $E$ connecting one vertex of $S$ with one vertex of $T$.
We let $G (S)$ denote the subgraph of $G$ induced on the vertices in $S$:
  the graph with vertex set $S$ containing the edges of $E$ between
  vertices in $S$.

For $S \subseteq V$, we define $\vol{S} = \sum_{i \in S} d_{i}$.
Observe that $\vol{V} = 2m$ if $G$ has $m$ edges.
The conductance of a set of vertices $S$, written
  $\conduc{G}{S}$, is often defined by
\[
  \conduc{G}{S} \defeq
  \frac{\sizeof{E (S, V - S)}}
       {\min \left(\vol{S}, \vol{V - S} \right)}.
\]
The conductance of $G$ is then given by
\[
  \Conduc{G}{} \defeq \min_{\emptyset \not = S \subset V} \conduc{}{S}.
\]

The conductance of a graph is related  to the smallest non-zero
  eigenvalue of its Laplacian matrix,
  but is even more strongly related to the
  smallest non-zero eigenvalue of its Normalized Laplacian
  matrix (see~\cite{Chung}), whose definition we now recall.
Let $D$ be the diagonal matrix whose $v$-th diagonal is $d_{v}$.
The Normalized Laplacian of the graph $G$, written $\calL_{G}$,
  is defined by
\[
  \calL_{G} = D^{-1/2} L_{G} D^{-1/2}.
\]
It is well-known that both $L_{G}$ and $\calL_{G}$ are positive semi-definite matrices,
  with smallest eigenvalue zero.
The eigenvalue zero has multiplicity one if an only if the graph $G$ is connected, in which case
  the eigenvector of $L_{G}$ with eigenvalue zero is the constant vector
  (see~\cite[page 269]{BollobasMGT}, or derive from \eqref{eqn:quadraticForm}).

Our analysis exploits a discreet version of
  Cheeger's inequality\cite{Cheeger} (see~\cite{Chung,JerrumSinclair,DiaconisStrook}), which relates the smallest non-zero
  eigenvalue of $\calL_{G}$, written $\lambda_{2} (\calL_{G})$, to the conductance of $G$.
\begin{theorem}[Cheeger's Inequality]\label{thm:cheeger}
\[
2 \Phi_{G} \geq \lambda_{2} (\calL_{G}) \geq \Phi_{G}^{2}/2.
\]
\end{theorem}

\section{A few examples}\label{sec:examples}

\subsection{Example 1: Complete Graph}
\begin{figure}[h]
\centering
\subfigure[$G$: The complete graph on $10$ vertices]{\epsfig{figure=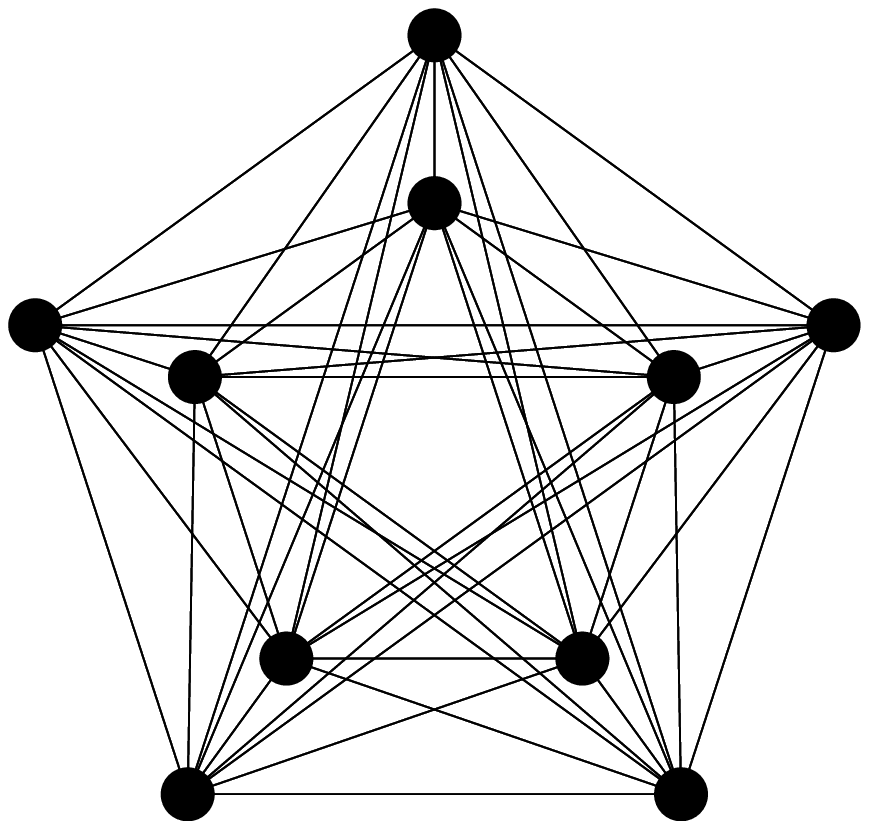,height=1.7in}}
\quad
\subfigure[$\Gtil$: A $\sqrt{5/2}$-approximation of $G$]{\epsfig{figure=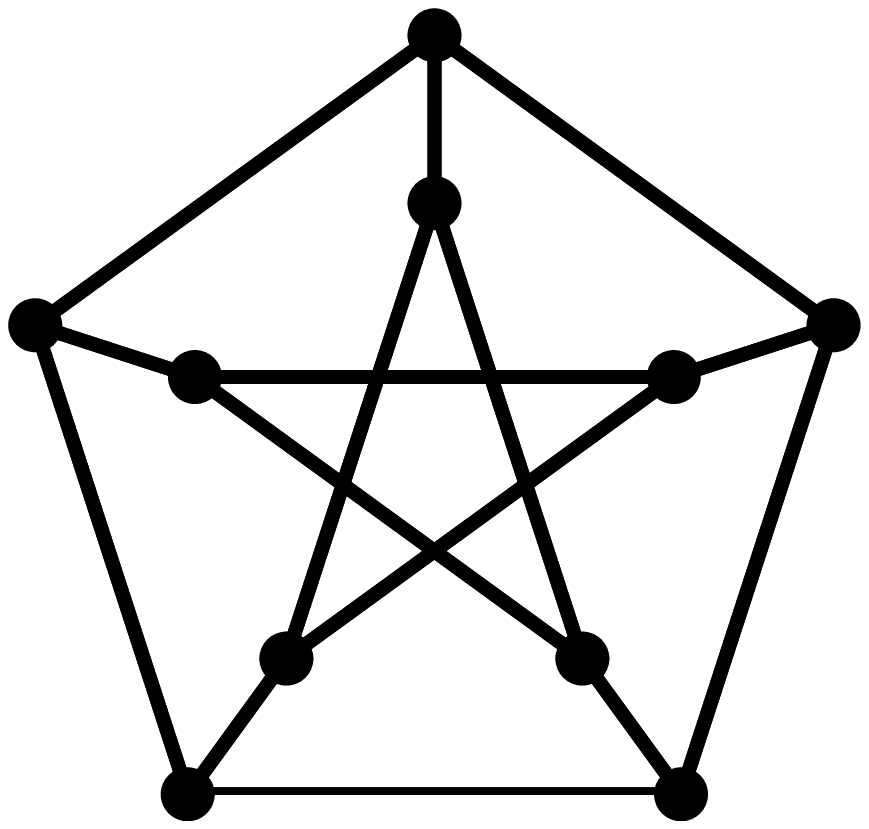,height=1.7in}}
\end{figure}

We first consider what a sparsifier of the complete graph should look like.
Let $G$ be the complete graph on $n$ vertices.
All non-zero eigenvalues of $L_{G}$ equal $n$.
So, for every unit vector $x$ orthogonal to the all-1s vector,
\[
  x^{T} L_{G} x = n.
\]
From Cheeger's inequality, one may prove that graphs with constant conductance, called expanders,
  have a similar property.
Spectrally speaking, the best of them
  are the Ramanujan graphs~\cite{LPS,Margulis}, which are $d$-regular
  graphs
  all of whose non-zero Laplacian eigenvalues lie between 
  $d-2\sqrt{d-1}$ and $d+2\sqrt{d+1}$.
So, if we let $\Gtil$ be a Ramanujan graph in which every edge has been given weight $n/d$,
  then for every unit vector $x$ orthogonal to the all-1s vector,
\[
x^{T} L_{\Gtil} x \in
 \left[ n- \frac{2 n \sqrt{d-1}}{d},
        n+ \frac{2 n \sqrt{d-1}}{d}
\right].
\]
Thus, $\Gtil$ is a $\left(1-2 \sqrt{d-1}/d \right)^{-1}$-approximation of $G$.

\subsection{Example 2: Joined Complete Graphs}

\begin{figure}[h]
\centering
\subfigure[$G$: Two complete graphs joined by an edge.]{\epsfig{figure=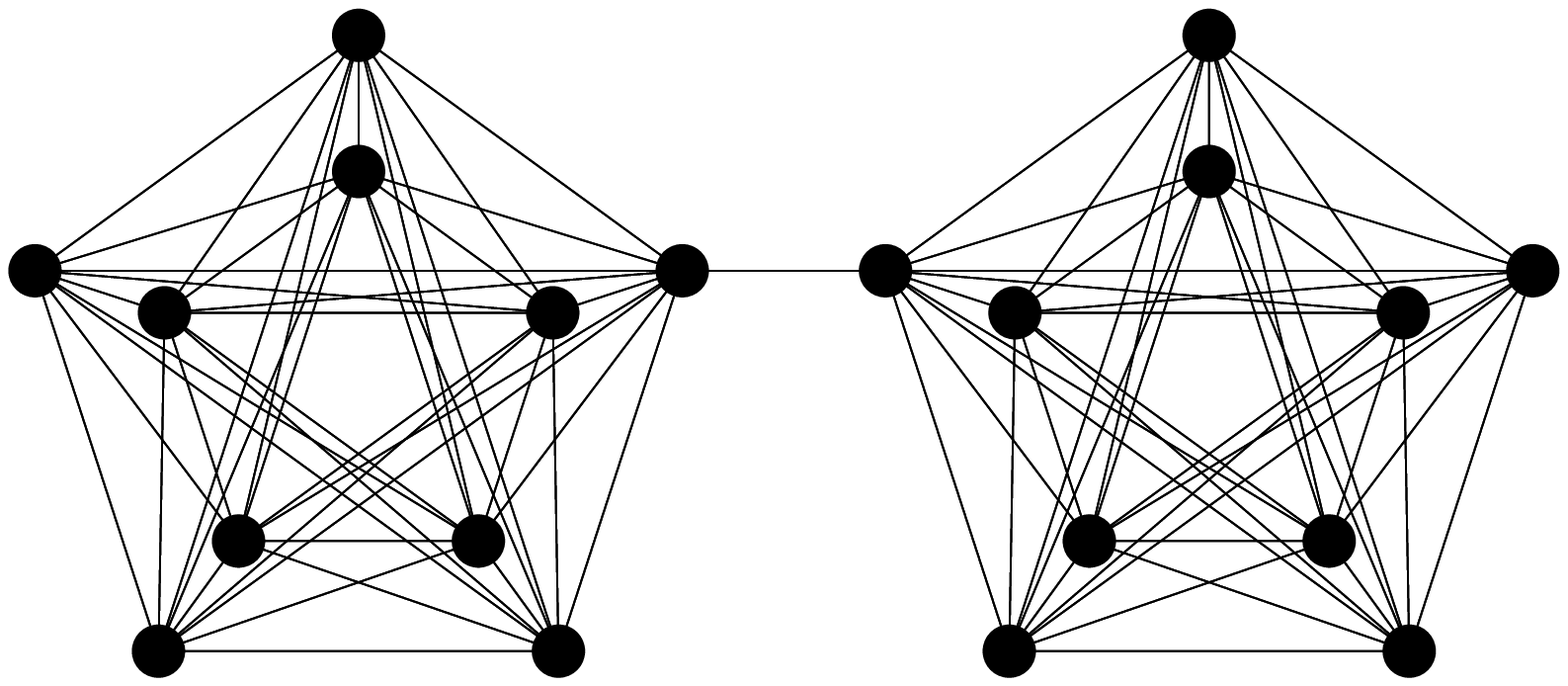,height=1.2in}} \quad
\subfigure[$\Gtil$: A good approximation of $G$.  Thicker edges indicate edges of weight $3$]{\epsfig{figure=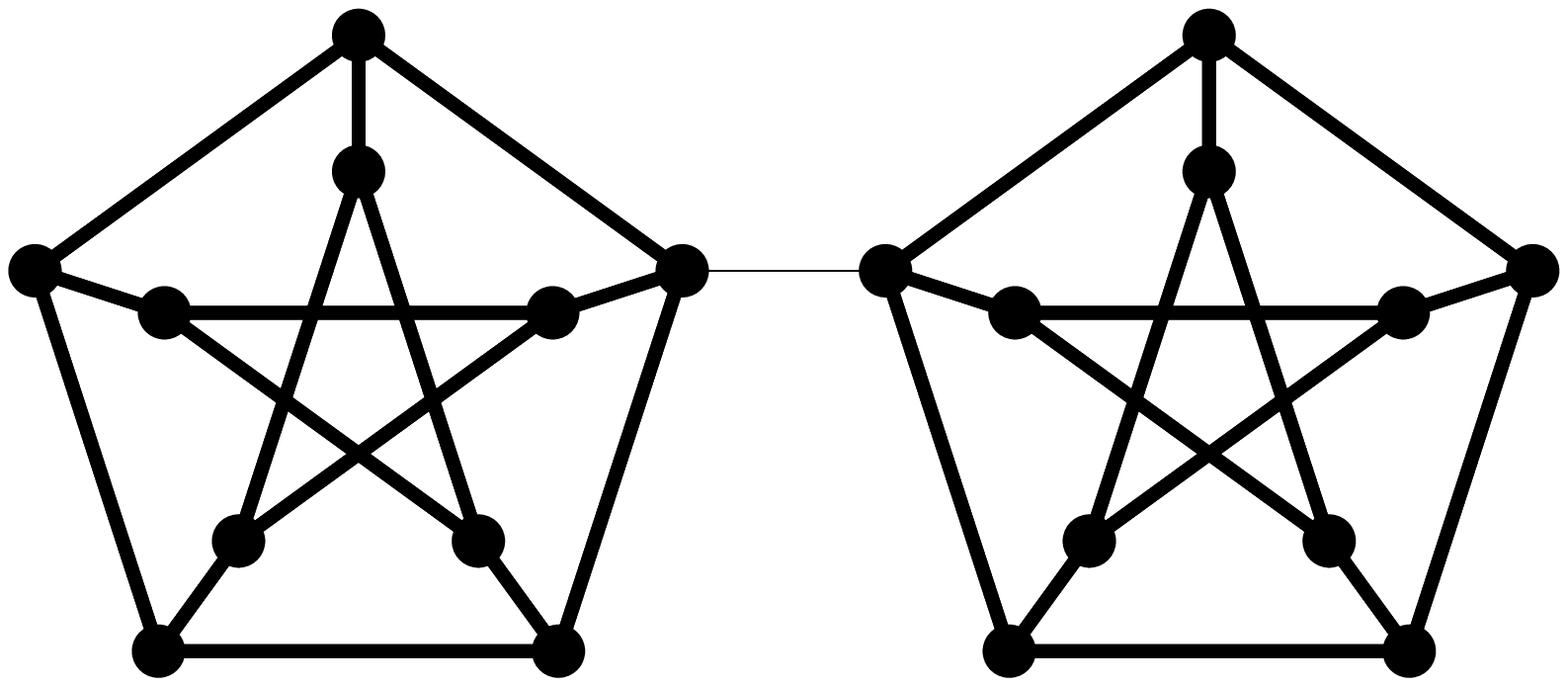,height=1.2in}}
\end{figure}

Next, consider a graph on $2n$ vertices obtained by joining two complete graphs
  on $n$ vertices
  by a single edge, $e$.
Let $V_{1}$ and $V_{2}$ be the vertex sets of the two complete graphs.
We claim that a good sparsifier for $G$ may be obtained by setting $\Gtil$ to be
  the edge $e$ with weight 1, plus $(n/d)$ times a Ramanujan graph on each vertex set.
To prove this, let $G_{1}$ and $G_{2}$ denote the complete graphs on $V_{1}$ and $V_{2}$,
  and let $G_{3}$ denote the graph just consisting of the edge $e$.
Similarly, let $\Gtil_{1}$ and $\Gtil_{2}$ denote $(n/d)$ times a Ramanujan graph on each
  vertex set, and let $\Gtil_{3} = G_{3}$.
Recalling the addition we defined on graphs, we have
\begin{align*}
  G & = G_{1} + G_{2} + G_{3}, \quad  \text{and} \\
  \Gtil & = \Gtil_{1} + \Gtil_{2} + \Gtil_{3}.
\end{align*}
We already know that for
$\sigma = \left(1-2 \sqrt{d-1}/d \right)^{-1}$,
and $i \in \setof{1,2}$ 
\[
  \frac{1}{\sigma} \Gtil_{i} \pleq G_{i} \pleq \sigma \Gtil_{i}.
\]
As $\Gtil_{3} = G_{3}$, we have
\[
  G  = G_{1} + G_{2} + G_{3}
  \pleq  \sigma \Gtil_{1} + \sigma \Gtil_{2} + \Gtil_{3}
  \pleq  \sigma \Gtil_{1} + \sigma \Gtil_{2} + \sigma \Gtil_{3}
  = \sigma \Gtil.
\]
The other inequality follows by similar reasoning.
This example demonstrates both the utility of using edges with different weights, even
  when sparsifying unweighted graphs, and how we can combine sparsifiers
  of subgraphs to sparsify an entire graph.
Also observe that every sparsifier of $G$ must contain the edge $e$, while no other
  edge is particularly important.

\subsection{Example 3: Distinguishing cut sparsifiers from spectral sparsifiers}

\begin{figure}[h]
\centering
\subfigure[$G$: $n=8$ sets of $k = 4$ vertices arranged in a ring and
  connected by complete bipartite graphs, plus one edge across.]{\epsfig{figure=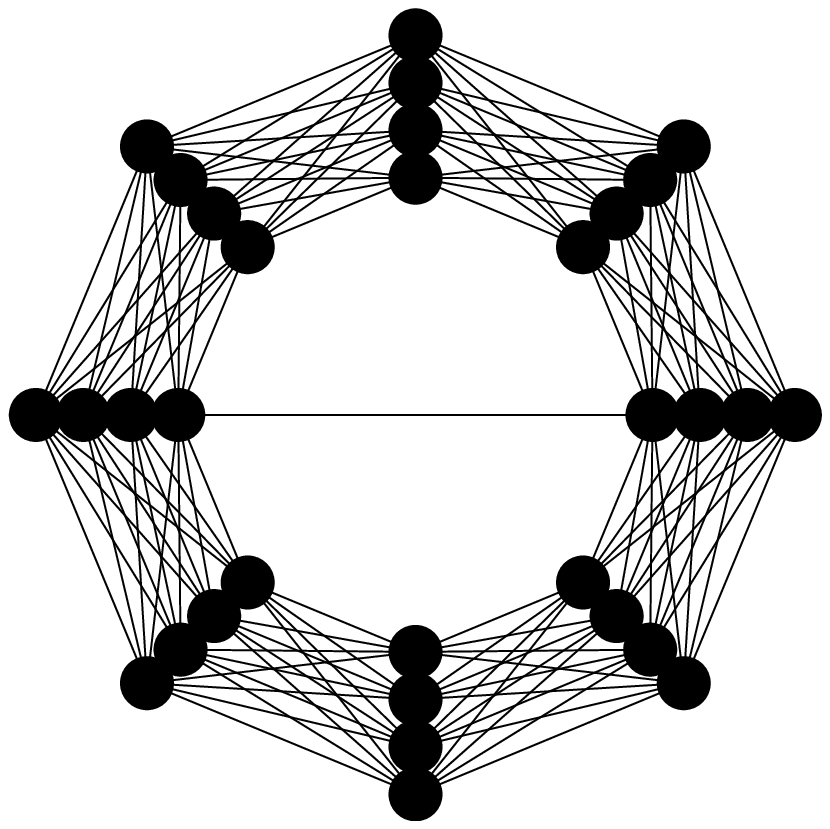,height=2in}} \quad
\subfigure[$\Gtil$: A good cut sparsifier of $G$, but a poor spectral sparsifier]{\epsfig{figure=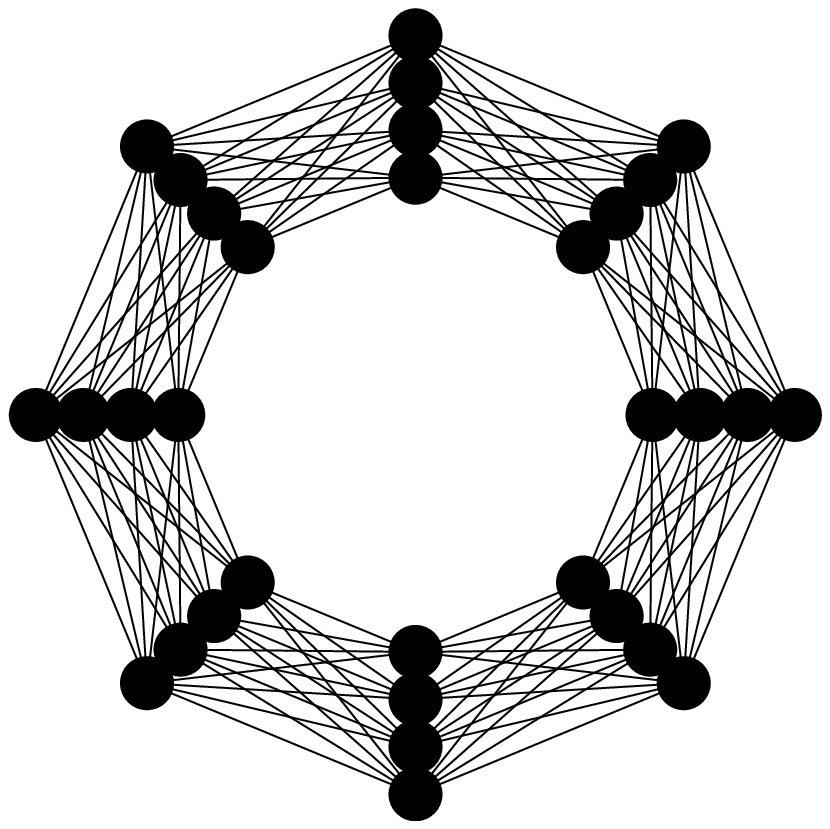,height=2in}}
\end{figure}


Our last example will demonstrate the difference between our notion of sparsification
  and that of Benczur and Karger.
We will describe graphs $G$ and $\Gtil$ for which $\Gtil$ is not a $\sigma$-approximation of $G$ for any small
  $\sigma$, but it is a very good sparsifier of $G$
  under the definition considered by Benczur and Karger.
The vertex set $V$ will be $\setof{0,\dotsc ,n-1} \times \setof{1, \dotsc , k}$, where $n$ is even.
The graph $\Gtil$ will consist of $n$ complete bipartite graphs, connecting all pairs of
  vertices $(u,i)$ and $(v,j)$ where $v = u \pm 1\mod n$.
The graph $G$ will be identical to the graph $\Gtil$, except that it will have one additional
  edge $e$ from vertex $(0,1)$ to vertex $(n/2,1)$.
As the minimum cut of $G$ has size $2k$, and $\Gtil$ only differs by one edge,
  $\Gtil$ is a $(1+1/2k)$-approximation of $G$ in the notion considered by Benczur and Karger.
To show that $\Gtil$ is a poor spectral approximation of $G$, consider the vector $x$
  given by
\[
x (u,i) = \min (u, n - u).
\]
One can verify that
\[
  x^{T} L_{\Gtil} x = n k^{2}, \quad \text{while} \quad
  x^{T} L_{G} x = n k^{2} + (n/2)^{2}.
\]
So,  inequality \eqref{eqn:approxForm}
  is not satisfied for any $\sigma$ less than $1 + n / 4 k^{2}$.

\section{Sampling Graphs}\label{sec:sampling}
In this section, we show that if a graph has high conductance,
  then it may be sparsified by a simple random sampling procedure.
The sampling procedure involves assigning a probability
  $p_{i,j}$ to each edge $(i,j)$, and then selecting edge $(i,j)$ to be
  in the graph $\Gtil$ with probability $p_{i,j}$.
When edge $(i,j)$ is chosen to be in the graph, we multiply its weight by
  $1/p_{i,j}$.
As the graph is undirected, we implicitly assume that $p_{i,j} = p_{j,i}$.
Let $A$ denote the adjacency matrix of the original graph $G$, and $\Atil$
  the adjacency matrix of the sampled graph $\Gtil$.
This procedure guarantees that
\[
  \expec{}{\Atil} = A.
\]
Sampling procedures of this form were examined by Benczur and Karger~\cite{BenczurKarger}
  and Achlioptas and McSherry~\cite{AchlioptasMcSherry}.
Achlioptas and McSherry analyze the approximation obtained by such a procedure
  through a bound on the norm of a random matrix
  of \Furedi \ and \Komlos~\cite{FurediKomlos}.
As their bound does not suffice for our purposes, we tighten it by
  refining the analysis of \Furedi \  and \Komlos.

If $\Gtil$ is going to be a sparsifier for $G$, then we must be sure
  that every vertex in $\Gtil$ has edges attached to it.
We guarantee this by requiring that, for some parameter $\Upsilon > 1$,
\begin{equation}\label{eqn:delta}
  p_{i,j}
  =
\min \left(1, \frac{\Upsilon}{\min (d_{i}, d_{j})} \right) , \quad
\text{for all edges $(i,j)$.}
\end{equation}
The parameter $\Upsilon$ controls the number of edges we expect to find
  in the graph, and will be set to at least $\Omega\left(\log n \right)$
  to ensure that every vertex has an attached edge.

We will show that if $G$ has high conductance and
 \eqref{eqn:delta} is satisfied for a sufficiently large $\Upsilon$,
 then $\Gtil$ will be a good sparsifier of $G$ with high probability.
The actual theorem that we prove is slightly more complicated,
  as it considers the case where we only apply the sampling
  on a subgraph of $G$.

\begin{theorem}[Sampling High-Conductance Graphs]\label{thm:sampling}
Let $\epsilon ,p \in (0,1/2)$ and
  let $G = (V,E)$ be an unweighted graph  whose smallest non-zero normalized Laplacian
  eigenvalue is at least $\lambda$.
Let $S$ be a subset of the vertices of $G$, let $F$ be the edges in $G (S)$,
  and let $H = E - F$ be the rest of the edges.
Let
\[
  (S, \Ftil) = \sample ((S,F), \epsilon , p, \lambda),
\]
and let $\Gtil = (V, \Ftil \union H)$.
Then, with probability at least $1-p$,
\begin{itemize}
\item [(S.1)] $\Gtil$ is a $(1+\epsilon)$-approximation of $G$, and
\item [(S.2)] The number of edges in $\Ftil$ is at most
\[
\frac{288 \max \left(\log_{2} (3/p), \log_{2} n \right)^{2}}
  {(\epsilon \lambda)^{2}}  \sizeof{S}.
\]
\end{itemize}
\end{theorem}

\vskip 0.2in
\noindent
\fbox{
\begin{minipage}{6in}
\noindent $\Gtil = \sample  (G, \epsilon , p, \lambda )$
\begin{enumerate}
\item [1.] Set $k = \max \left(\log_{2} (3/p), \log_{2} n \right)$.
\item [2.] Set $\Upsilon = \left(\frac{12 k}{\epsilon \lambda} \right)^{2}$.
\item [3.] For every edge $(i,j)$ in $G$, set
  $p_{i,j} = \min \left(1, \frac{\Upsilon}{\min (d_{i}, d_{j})} \right)$.
\item [4.] For every edge $(i,j)$ in $G$, with probability
  $p_{i,j}$ put an edge of weight $1/p_{i,j}$ between vertices $(i,j)$
  into $\Gtil$.
\end{enumerate}
\end{minipage}
}
\vskip 0.2in

Let $D$ be the diagonal matrix of degrees of vertices of $G$.
To prove Theorem~\ref{thm:sampling}, we  establish that
  the 2-norm of $D^{-1/2} (L_{G} - L_{\Gtil }) D^{-1/2}$ is probably small\footnote{%
Recall that the 2-norm of a symmetric matrix is the largest absolute value of
its eigenvalues.}, and then apply the following lemma.

\begin{lemma}\label{lem:normSmallApprox}
Let $L$ be the Laplacian matrix of a connected graph $G$, $\Ltil$ be the Laplacian of $\Gtil$,
  and let $D$ be the diagonal matrix of degrees of $G$.
If
\begin{itemize}
\item [1.] $\lambda_{2} \left(D^{-1/2} L D^{-1/2} \right) \geq \lambda $, and
\item [2.] $\norm{D^{-1/2} (L - \Ltil)D^{-1/2}} \leq \epsilon$,
\end{itemize}
then $\Gtil$ is a $\sigma$-approximation of $G$ for
\[
  \sigma = \frac{\lambda}{\lambda - \epsilon}.
\]
\end{lemma}
\begin{proof}
Let $x$ be any vector and let $y = D^{1/2} x$.
By assumption, $G$ is connected and so
  the nullspace of the normalized Laplacian $D^{-1/2} L D^{-1/2}$
  is spanned by $D^{1/2} \bvec{1}$.
Let $z$ be the projection of $y$ orthogonal to
  $D^{1/2} \bvec{1}$, so
\begin{equation}\label{eqn:normSmallApprox1}
  x^{T} L x =
  y^{T} D^{-1/2} L D^{-1/2} y =
  z^{T} \left(D^{-1/2} L D^{-1/2} \right) z \geq \lambda \norm{z}^{2}.
\end{equation}
We compute
\begin{align*}
  x^{T} \Ltil x
 & =
  y^{T} D^{-1/2} \Ltil D^{-1/2} y \\
 & =
  z^{T} D^{-1/2} \Ltil D^{-1/2} z \\
 & =
  z^{T} D^{-1/2} L D^{-1/2} z
+
  z^{T} D^{-1/2} (\Ltil-L) D^{-1/2} z \\
 & =
  z^{T} D^{-1/2} L D^{-1/2} z
\left(
1 +
\frac{z^{T} D^{-1/2} (\Ltil-L) D^{-1/2} z}
     {  z^{T} D^{-1/2} L D^{-1/2} z }
 \right)
   \\
 & \geq
  z^{T} D^{-1/2} L D^{-1/2} z
\left(
1 -
\frac{\epsilon \norm{z}^{2}}
     {\lambda \norm{z}^{2}}
\right)
\quad
\text{(by assumption 2 and \eqref{eqn:normSmallApprox1})}
\\
& = \left(\frac{\lambda -\epsilon}{\lambda} \right) x^{T} L x.
\quad \text{(again by \eqref{eqn:normSmallApprox1})}
\end{align*}
We may similarly show that
\[
x^{T} \Ltil x \leq \left(\frac{\lambda +\epsilon}{\lambda} \right) x^{T} L x \leq \left(\frac{\lambda}{\lambda -\epsilon} \right) x^{T} L x.
\]
The lemma follows from these inequalities.
\end{proof}

Let $A$ be the adjacency matrix of $G$ and let $\Atil$ be the adjacency matrix
  of $\Gtil$.
For each edge $(i,j)$,
\[
  \Atil_{i,j} =
\begin{cases}
1 / p_{i,j} &
\text{ with probability $p_{i,j}$ and}
\\
0  & \text{ with probability $1-p_{i,j}$.}
\end{cases}
\]

To prove Theorem~\ref{thm:sampling}, we will observe that
\[
\norm{D^{-1/2} (L - \Ltil) D^{-1/2}}
\leq
\norm{D^{-1/2} (A - \Atil) D^{-1/2}}
+
\norm{D^{-1/2} (D - \Dtil) D^{-1/2}},
\]
where $\Dtil$ is the diagonal matrix of the diagonal entries of $\Ltil$.
It will be easy to bound the second of these terms, so we defer that part of the proof
  to the end of the section.
A bound on the first term comes from the following lemma.

\begin{lemma}[Random Subgraph]\label{lem:sampling}
For all even integers $k$,
\[
  \prob{}{\norm{D^{-1/2} (\Atil  - A) D^{-1/2}} \geq
 \frac{2 k n^{1/k} }
      {\sqrt{\Upsilon }}
         }
  \leq  2 ^{-k}.
\]
\end{lemma}

Our proof of this lemma applies a modification of techniques introduced
  by \Furedi \ and \Komlos~\cite{FurediKomlos} (See also the paper by
  Vu~\cite{Vu} that corrects some bugs in their work).
However, they consider the eigenvalues of random graphs in which
  every edge can appear.
Some interesting modifications are required to make an argument such
  as ours work when downsampling a graph that may already be sparse.
We remark that without too much work one can
  generalize Theorem~\ref{thm:sampling} so that it applies
  to weighted graphs.

\begin{proof}[Proof of Lemma~\ref{lem:sampling}]
To simplify notation, define
\[
\Delta  = D^{-1} (\Atil - A),
\]
so for each edge $(i,j)$,
\[
\Delta_{i,j} = \begin{cases}
\frac{1}{d_{i}} (\frac{1}{p_{i,j}} - 1)  &
 \text{ with probability $p_{i,j}$, and}
\\
- \frac{1}{d_{i}}
  &
 \text{ with probability $1- p_{i,j}$.}
\end{cases}
\]

Note that $D^{-1/2} (\Atil - A) D^{-1/2}$ has the same eigenvalues
  as $\Delta$.
So, it suffices to bound the absolute values of the eigenvalues of $\Delta$.
Rather than trying to upper bound the eigenvalues of $\Delta $
  directly, we will upper bound a power of $\Delta$'s trace.
As the trace of a matrix is the sum of its eigenvalues,
  $\trace{\Delta ^{k}}$ is an upper bound
  on the $k$th power of every eigenvalue of $\Delta$,
 for every even power $k$.

Lemma~\ref{lem:trace} implies that, for even $k$,
\[
\frac{n k^{k}}{\Upsilon^{k/2}}
\geq
\expec{}{\trace{\Delta^{k} }}
\geq
\expec{}{\lambda _{max}\left(\Delta^{k} \right)}.
\]
Applying Markov's inequality, we obtain
\[
\prob{}{\trace{\Delta^{k}} \geq  2^{k} \frac{n k^{k}}{\Upsilon^{k/2}}}
\leq
1/2^{k}.
\]
Recalling that
  the eigenvalues of $\Delta^{k}$ are the $k$-th powers of the eigenvalues of $\Delta$,
 and
  taking $k$-th roots, we conclude
\[
\prob{}{\norm{D^{-1/2} (\Atil  - A) D^{-1/2}} \geq 
  2 \frac{n^{1/k} k}{\Upsilon^{1/2}}}
\leq
1/2^{k}.
\]
\end{proof}

\begin{lemma}\label{lem:trace}
For even $k$,
\[
\expec{}{\trace{\Delta^{k}}}
\leq
\frac{n k^{k}}{\Upsilon^{k/2}}.
\]
\end{lemma}
\begin{proof}
Recall that the $(v_{0}, v_{k})$ entry of $\Delta^{k}$ satisfies
\[
\left(\Delta^{k} \right)_{v_{0}, v_{k}}
=
\sum_{\vs{v}{1}{k-1}} \prod_{i=1}^{k} \Delta_{v_{i-1}, v_{i}}.
\]
Taking expectations, we obtain
\begin{equation}\label{eqn:deltaExpec}
\expec{}{\left(\Delta^{k} \right)_{v_{0}, v_{k}}}
=
\sum_{\vs{v}{1}{k-1}}
\expec{}{\prod_{i=1}^{k} \Delta_{v_{i-1}, v_{i}}}.
\end{equation}
We will now describe a way of coding every sequence
  $v_{1}, \dotsc , v_{k-1}$
  that could
  possibly contribute to the sum.
Of course, any sequence containing a consecutive pair
  $(v_{i-1}, v_{i})$ for
  which $\Delta _{v_{i-1}, v_{i}}$ is always zero
  will contribute zero to the sum.
So, for a sequence to have a non-zero contribution, each
  consecutive pair $(v_{i-1}, v_{i})$ must be an edge in the graph $A$.
Thus, we can identify every sequence with non-zero contribution with a walk
  on the graph $A$ from vertex $v_{0}$ to vertex $v_{k}$.

The first idea in our analysis is to observe that most of the
  terms in this sum are zero.
The reason is that, for all $v_{i}$ and $v_{j}$
\[
\expec{}{\Delta_{v_{i},v_{j}}}
 = 0.
\]
As $\Delta _{v_{i}, v_{j}}$ is independent of every
  term in $\Delta $ other than $\Delta _{v_{j}, v_{i}}$,
  we see that the term
\begin{equation}\label{eqn:deltaTerm}
\expec{}{\prod_{i=1}^{k} \Delta_{v_{i-1}, v_{i}}},
\end{equation}
corresponding to
  $v_{1}, \dotsc , v_{k-1}$, will be zero
  unless each edge $(v_{i-1}, v_{i})$ appears at least
  twice (in either direction).

We now describe a method for coding all walks
  in which each edges appears at least twice.
We set $T$ to be the set of time steps $i$
  at which the edge between $v_{i-1}$ and $v_{i}$
  does not appear earlier in the walk (in either direction).
Note that $1$ is always an element of $T$.
We then let $\tau $ denote the map from
  $[k] - T \rightarrow T$, indicating for each
  time step not in $T$ the time step in which
  the edge traversed first appeared (regardless of
  in which direction it is traversed).
Note that we need only consider the cases in which $\sizeof{T} \leq k/2$, as
  otherwise some edge appears only once in the walk.
To finish our description of a walk, we need
  a map
\[
  \sigma : T \rightarrow \setof{1,\dotsc ,n},
\]
indicating the vertex encountered at each time $i \in T$.

For example, for the walk
\[
\begin{tabular}{| l || l | l |  l | l |  l | l |  l | l |  l | l | l |}
\hline
Step & 0 & 1 & 2 & 3 & 4 & 5 & 6 &7 & 8 & 9 & 10\\
Vertex & a & b & c & d & b & c & d &b & e & b & a\\
\hline
\end{tabular},
\]
we get
\[
  T = \setof{1, 2, 3, 4, 8}
\qquad
\tau :
\begin{aligned}
5 & \mapsto 2\\
6 & \mapsto 3\\
7 & \mapsto 4\\
9 & \mapsto 8\\
10 & \mapsto 1\\
\end{aligned}
\qquad
\sigma  :
\begin{aligned}
1 & \mapsto b\\
2 & \mapsto c\\
3 & \mapsto d\\
4 & \mapsto b\\
8 & \mapsto e\\
\end{aligned}
\]

Using $T$, $\tau$ and $\sigma$, we can inductively reconstruct
  the sequence $v_{1}, \dotsc , v_{k-1}$ by the rules
\begin{itemize}
\item if $i \in T$, $v_{i} = \sigma (i)$,
\item if $i \not \in T$, and $v_{i-1} = v_{\tau (i) -1}$,
  then $v_{i} = v_{\tau (i)}$, and
\item if $i \not \in T$, and $v_{i-1} = v_{\tau (i)}$,
  then $v_{i} = v_{\tau (i)-1}$.
\end{itemize}
If $v_{i-1} \not \in \setof{v_{\tau (i)}, v_{\tau (i)-1}}$,
  then the tuple $(T, \tau , \sigma)$ does not properly code
  a walk on the graph of $A$.
We will call $\sigma $ a \textit{valid assignment} for $T$ and $\tau $ if
  the above rules do produce a walk on the graph of $A$ from $v_{0}$
  to $v_{k}$.

We have
\begin{align}
\expec{}{  \left(\Delta ^{k} \right)_{v_{0}, v_{k}}}
& =
  \sum _{T, \tau }
  \sum _{\text{$\sigma$ valid for $T$ and $\tau $} }
  \expec{}{
    \prod _{i=1}^{k} \Delta _{v_{i-1},v_{i}}
  }, \notag
\intertext{(where $(\vs{v}{1}{k-1})$ is the sequence encoded by $(T, \tau , \sigma)$)}
\notag
\\
& =
  \sum _{T, \tau }
  \sum _{\text{$\sigma$ valid for $T$ and $\tau $} }
  \prod _{s \in T}
  \expec{}{
    \Delta _{v_{s-1},v_{s}}
    \prod _{i : \tau(i) = s}
    \Delta _{v_{i-1},v_{i}}
  }. \label{eqn:sampling1}
\end{align}

Each of the terms
\[
  \expec{}{
    \Delta _{v_{s-1},v_{s}}
    \prod _{i : \tau(i) = s}
    \Delta _{v_{i-1},v_{i}}
}
\]
is independent of the others,
  and involves a product of the terms
  $\Delta _{v_{s-1},v_{s}}$
  and
  $\Delta _{v_{s},v_{s-1}}$.
In Lemma~\ref{lem:edgeExpec}, we will prove that
\begin{equation}\label{eqn:simpleSample}
    \expec{}{
    \Delta _{v_{s-1},v_{s}}
    \prod _{i : \tau(i) = s}
    \Delta _{v_{i-1},v_{i}}
  }
 \leq
   \frac{1}{\Upsilon ^{\sizeof{\setof{i : \tau (i) = s}}}}
  \frac{1}{d_{v_{s-1}}},
\end{equation}
which implies
\begin{equation}\label{eqn:simpleSample2}
  \sum _{\text{$\sigma$ valid for $T$ and $\tau $} }
  \prod _{s \in T}
  \expec{}{
    \Delta _{v_{s-1},v_{s}}
    \prod _{i : \tau(i) = s}
    \Delta _{v_{i-1},v_{i}}
  }
\leq
  \frac{1}{\Upsilon ^{k-\sizeof{T}}}
  \sum _{\text{$\sigma$ valid for $T$ and $\tau $} }
  \prod _{s \in T}
  \frac{1}{d_{v_{s-1}}}.
\end{equation}
To bound the sum of products on the right hand-side of
  \eqref{eqn:simpleSample2}, fix $T$ and $\tau$ and consider
  the following random process for generating a valid $\sigma$
  and corresponding walk:
  go through the elements of $T$ in order.
For each $s \in T$,
  pick $\sigma (s)$ to be a random neighbor of the $s-1$st vertex in the walk.
If possible, continue the walk according to $\tau$ until it reaches the next
  step in $T$.
If the process produces a valid $\sigma$, return it.
Otherwise, return nothing.
The probability that any particular valid $\sigma$ will be returned by this
  process is
\[
  \prod_{s \in T} \frac{1}{d_{v_{s-1}}}.
\]
So,
\begin{equation}\label{eqn:simpleSample3}
  \sum _{\text{$\sigma$ valid for $T$ and $\tau $} }
  \prod _{s \in T}
  \frac{1}{d_{v_{s-1}}}
\leq
  1.
\end{equation}

As there are at most
  at most $2^{k}$ choices for $T$,
  and at most $\sizeof{T}^{k-\sizeof{T}} \leq \sizeof{T}^k$ choices for $\tau $,
  we may combine inequalities \eqref{eqn:simpleSample2} and \eqref{eqn:simpleSample3} 
  with
  \eqref{eqn:sampling1} to obtain
\[
\expec{}{  \left(\Delta ^{k} \right)_{v_{0}, v_{k}}}
 \leq \frac{(2\sizeof{T})^{k}}{\Upsilon ^{k-\sizeof{T}}} \leq
  \frac{k^{k}}{\Upsilon ^{k/2}}. \quad \text{(using $\sizeof{T} \leq k/2$)}
\]
The lemma now follows from
\[
\expec{}{\trace{\Delta ^{k}}}
=
\sum_{v_{0}=1}^{n}\expec{}{  \left(\Delta ^{k} \right)_{v_{0}, v_{0}}}.
\]
\end{proof}

\begin{claim}\label{clm:magnitude}
\[
\abs{\Delta_{i,j}}
\leq 1/\Upsilon .
\]
\end{claim}
\begin{proof}
If $p_{i,j} = 1$, then $\Delta_{i,j} = 0$.
If not, then we have $\Upsilon / \min (d_{i}, d_{j}) = p_{i,j} < 1$.
With probability $1-p_{i,j}$,
\[
\abs{\Delta_{i,j}} = \frac{1}{d_{i}} \leq \frac{1}{\min (d_{i}, d_{j})}
 \leq  1/ \Upsilon.
\]
On the other hand, with probability $p_{i,j}$,
\[
\Delta_{i,j} = \frac{1}{d_{i}}\left(\frac{1}{p_{i,j}} - 1 \right)
\leq \frac{1}{d_{i}} \frac{1}{p_{i,j}}
\leq \frac{1}{\min (d_{i}, d_{j})} \frac{1}{p_{i,j}}
= 1/\Upsilon.
\]
As $\Delta_{i,j} \geq 0$ in this case, we have established
  $\abs{\Delta_{i,j}} \leq 1/\Upsilon$.
\end{proof}

\begin{lemma}\label{lem:edgeExpec}
For all edges $(r,t)$ and integers $k \geq 1$ and $l \geq 0$,
\[
  \expec{}{\Delta_{r,t}^{k} \Delta_{t,r}^{l}}
\leq
\frac{1}{\Upsilon ^{k+l - 1}}
\frac{1}{d_{r}}.
\]
\end{lemma}
\begin{proof}
First, if $p_{i,j} = 1$, then $\Delta_{i,j} = 0$.
Second, if $k + l = 1$, $\expec{}{\Delta_{r,t}^{k} \Delta_{t,r}^{l}} = 0$.
So, we may restrict our attention to the case where $k + l \geq 2$ and $p_{i,j} < 1$,
  which by \eqref{eqn:delta} implies $p_{i,j} = \Upsilon / \min (d_{r}, d_{t})$.
Claim~\ref{clm:magnitude} tells us that for $k \geq 1$,
\[
  \expec{}{\Delta_{r,t}^{k} \Delta_{t,r}^{l}}
\leq
\frac{1}{\Upsilon }
  \expec{}{\Delta_{r,t}^{k-1} \Delta_{t,r}^{l}}.
\]
A similar statement may be made for $l \geq 1$.
So, it suffices to prove the lemma in the case
  $k + l = 2$.

As $\Delta_{r,t} = (\Atil_{r,t} -1 )/ d_{r}$ and
  $\Delta_{t,r} = (\Atil_{r,t} -1 )/ d_{t}$, we have
\begin{align*}
\expec{}{\Delta_{r,t}^{k} \Delta_{t,r}^{l}}
& =
\frac{1}{d_{r}^{k} d_{t}^{l}}
\expec{}{(\Atil_{r,t} -1 )^{k+l}}
\\
& =
\frac{1}{d_{r}^{k} d_{t}^{l}}
\left(
p_{r,t}
\left(\frac{1 - p_{r,t}}{p_{r,t}} \right)^{2}
+
(1- p_{r,t})
 \right) & (\text{using $k + l = 2$})\\
& =
\frac{1}{d_{r}^{k} d_{t}^{l}}
\left(\frac{1 - p_{r,t}}{p_{r,t}} \right)
\\
& \leq
\frac{1}{d_{r}^{k} d_{t}^{l}}
\left(\frac{1}{p_{r,t}} \right)
\\
& =
\frac{1}{d_{r}^{k} d_{t}^{l}}
\left(\frac{\min (d_{r}, d_{t})}{\Upsilon} \right).
\end{align*}
In the case $k = 1$, $l = 1$, we finish the proof by
\[
\frac{\min (d_{r}, d_{t})}{d_{r} d_{t}}
 =
\frac{1}{\max (d_{r}, d_{t})}
\leq
\frac{1}{d_{r}},
\]
and in the case $k = 2$, $l = 0$ by
\[
\frac{\min (d_{r}, d_{t})}{d_{r}^{2}}
\leq
\frac{1}{d_{r}}.
\]
\end{proof}

This finishes the proofs of Lemmas~\ref{lem:trace} and~\ref{lem:sampling}.
We now turn to the last ingredient we will need for the proof of Theorem~\ref{thm:sampling},
  a bound on the norm of the difference of the degree matrices.

\begin{lemma}\label{lem:sampledDegrees}
Let $G$ be a graph and let $\Gtil$ be obtained by sampling $G$ with probabilities
  $p_{i,j}$ that satisfy \eqref{eqn:delta}.
Let $D$ be the diagonal matrix
  of degrees of $G$, and let $\Dtil$ be the diagonal matrix of weighed degrees
  of $\Gtil$.
Then,
\[
  \prob{}{\norm{D^{-1/2} (D - \Dtil) D^{-1/2}} \geq \epsilon  }
  \leq 2 n e^{- \Upsilon \epsilon^{2} / 3}.
\]
\end{lemma}
\begin{proof}
Let $\dtil_{i}$ be the weighted degree of vertex $i$ in $\Gtil$.
As $D$ and $\Dtil$ are diagonal matrices,
\[
\norm{D^{-1/2} (D - \Dtil) D^{-1/2}}
=
\max_{i} \abs{1 - \frac{\dtil_{i}}{d_{i}}}.
\]
As the expectation of $\dtil_{i}$ is $d_{i}$ and
  $\dtil_{i}$ is a sum of $d_{i}$ random variables
  each of which is always $0$ or some value less than $d_{i}/ \Upsilon$,
  we may apply the variant of the Chernoff bound given in
  Theorem~\ref{thm:chernoff}
  to show that
\[
\prob{}{\abs{\dtil_{i} - d_{i}} > \epsilon d_{i}} \leq
2 e^{-\Upsilon \epsilon^{2} / 3}.
\]
The lemma now follows by taking a union bound over $i$.
\end{proof}

We use the following variant of the Chernoff bound from~\cite{Raghavan}.
\begin{theorem}[Chernoff Bound]\label{thm:chernoff}
Let $\alpha_{1}, \dotsc , \alpha_{n}$ all lie in $[0,\beta ]$ and 
 let $X_{1}, \dotsc , X_{n}$ be independent random variables such
  that $X_{i}$ equals $\alpha_{i}$ with probability $p_{i}$
  and $0$ with probability $1-p_{i}$.
Let $X = \sum_{i} X_{i}$ and
  $\mu = \expec{}{X} = \sum \alpha_{i} p_{i}$.
Then,
\[
\prob{}{X > (1+\epsilon) \mu}
 <
\left(\frac{e^{\epsilon}}{(1+\epsilon)^{1+\epsilon }} \right)^{\mu / \beta }
\quad
\text{and}
\quad 
\prob{}{X < (1-\epsilon) \mu}
 <
\left(\frac{e^{\epsilon}}{(1+\epsilon)^{1+\epsilon }} \right)^{\mu / \beta }
\]
For $\epsilon < 1$,
  both of these probabilities are at most
  $e^{-\mu \epsilon^{2}/3 \beta }$.
\end{theorem}
We remark that Raghavan~\cite{Raghavan} proved this theorem with $\beta = 1$;
  the extension to general $\beta > 0$ follows by re-scaling.

\begin{proof}[Proof of Theorem~\ref{thm:sampling}]
Let $L$ be the Laplacian of $G$, $A$ be its adjacency matrix, and $D$
  its diagonal matrix of degrees.
Let $\Ltil$, $\Atil$ and $\Dtil$ be the corresponding matrices for $\Gtil$.
The matrices $L$ and $\Ltil $ only differ on rows and columns indexed by $S$.
So, if we let $L (S)$ denote the submatrix of $L$ with rows and
  columns in $S$, we have
\begin{multline*}
  \norm{D^{-1/2} (L - \Ltil) D^{-1/2}}
=
  \norm{D (S)^{-1/2} (L (S) - \Ltil (S)) D (S)^{-1/2}}\\
\leq
  \norm{D (S)^{-1/2} (A (S) - \Atil (S)) D (S)^{-1/2}}
+
  \norm{D (S)^{-1/2} (D (S) - \Dtil (S)) D (S)^{-1/2}}.
\end{multline*}
Applying Lemma~\ref{lem:sampling} to the first of these terms,
  while observing
\[
  \frac{2 k n^{1/k}}{\sqrt{\Upsilon}}
\leq
  \frac{4 k }{\sqrt{\Upsilon}}
=
\frac{\epsilon \lambda}{3},
\]
  we find
\[
\prob{}{\norm{D (S)^{-1/2} (A (S) - \Atil (S)) D (S)^{-1/2}}
  \geq \frac{\epsilon \lambda}{3}} \leq p/3.
\]
Applying Lemma~\ref{lem:sampledDegrees} to the second term,
  we find
\[
\prob{}{\norm{D (S)^{-1/2} (D (S) - \Dtil (S)) D (S)^{-1/2}}
  \geq \frac{\epsilon \lambda}{3}}
 \leq
2 n e^{-\Upsilon \left(\epsilon \lambda /3 \right)^{2} / 3}
<
2 n e^{-2 k^{2}}
\leq
p/3.
\]
Thus, with probability at least $1-2p/3$,
\[
  \norm{D^{-1/2} (L - \Ltil) D^{-1/2}} \leq \frac{2 \epsilon \lambda}{3},
\]
in which case Lemma~\ref{lem:normSmallApprox} tells us that
  $\Gtil$ is a $\sigma$-approximation of $G$
  for
\[
  \sigma = \frac{\lambda}{\lambda - (2/3) \epsilon \lambda }
  \leq
  1+\epsilon,
\]
using
  $\epsilon \leq 1/2$.

Finally, we use Theorem~\ref{thm:chernoff} to bound the number of edges in $\Ftil$.
For each edge $(i,j)$ in $F$, let $X_{(i,j)}$ be the indicator random variable for
  the event that edge $(i,j)$ is chosen to appear in $\Ftil$.
Using $d_{i}$ to denote the degree of vertex $i$ in $G (S)$, we have
\begin{align*}
  \expec{}{\sum X_{(i,j)}}
& =
  \Upsilon \sum_{(i,j) \in F} \frac{1}{ \min (d_{i}, d_{j})}\\
& \leq
  \Upsilon \sum_{(i,j) \in F} \left(\frac{1}{d_{i}} + \frac{1}{d_{j}} \right)\\
& =
  \Upsilon \sum_{i \in S} \sum_{j : (i,j) \in F} \left(\frac{1}{d_{i}} \right)\\
& = \Upsilon \sizeof{S}.
\end{align*}
One may similarly show that $\expec{}{\sum X_{(i,j)}} \geq \Upsilon \sizeof{S} /2$.
Applying Theorem~\ref{thm:chernoff} with $\epsilon = 1$ (note that here $\epsilon$ is the
parameter in the statement of Theorem~\ref{thm:chernoff}), we obtain
\[
  \prob{}{\sum X_{(i,j)} \geq 2 \Upsilon \sizeof{S}}
\leq
\left(\frac{e}{4} \right)^{- \Upsilon \sizeof{S} / 2}
\leq
\left(\frac{e}{4} \right)^{- (8 \log_{2} (3/p))^{2}}
\leq p/3.
\]

\end{proof}

\section{Graph Decompositions}\label{sec:decomp}
In this section, we prove
  that every graph can be decomposed into components of high conductance,
  with a relatively small number of edges bridging the components.
A similar result was obtained independently by Trevisan~\cite{Trevisan}.
We prove this result for three reasons: first, it enables us to quickly establish
  the existence of good spectral sparsifiers.
Second, our algorithm for building sparsifiers requires a graph
  decomposition routine which is inspired by the computationally
  infeasible routine presented in this section%
\footnote{The routine \texttt{idealDecomp} is infeasible because it requires
  the solution of an NP-hard problem in step 2.
We could construct sparsifiers from a routine that approximately satisfies
  the guarantees of \texttt{idealDecomp}, such as the clustering algorithm of
  Kannan, Vempala and Vetta~\cite{KannanVempalaVetta}.
However, their routine could take  quadratic time, which is too slow
  for our purposes.
}.
Finally, the analysis of our algorithm relies upon Lemma~\ref{lem:certificate}, which
  occupies most of this section.
Throughout this section, we will consider an unweighted graph $G = (V,E)$,
  with $V = \setof{1,\dotsc ,n}$.
In the construction  of a decomposition of $G$, we will be concerned with
  vertex-induced subgraphs of $G$.
However, \textit{when measuring the conductance and volumes of vertices in
  these vertex-induced subgraphs, we will continue to measure the
  volume according to the degrees of vertices in the original graph.}
For clarity, 
  we define the boundary of a vertex set $S$ with respect to another
  vertex set $B$ to be
\[
  \bdry{B}{S} = E (S, B- S),
\]
we define the conductance of a set $S$ in the subgraph induced
  by $B \subseteq V$ to be
\[
  \conducin{B}{S} \defeq
  \frac{\sizeof{E (S, B - S)}}
       {\min \left(\vol{S}, \vol{B - S} \right)},
\]
and we define
\[
  \Conducin{B}
   \defeq \min_{S \subset B} \conducin{B}{S}.
\]
For convenience, we define $\conducin{B}{\emptyset} = 1$ and, for $\sizeof{B} = 1$,
  $\Conducin{B}{} = 1$.

We introduce the notation $G\{B \}$ to denote the graph $G (B)$ to which self-loops
  have been added so that every vertex in $G\{B \}$ has the same degree as in $G$.
For $S \subseteq B$
\[
  \conduc{G\{B \}}{S} =
 \conducin{B}{S}.
\]
Because $\Conducin{B}$ measures volume by degrees in $G$
  and those degrees are higher than in $G (B)$,
\[
  \Conducin{B}{} =  \Conduc{G\{B \}} \leq   \Conduc{G (B)}{}.
\]
So, when we prove lower bounds on $ \Conducin{B}{}$, we obtain lower
  bounds on $ \Conduc{G (B)}{}$.

\subsection{Spectral Decomposition}

We define a \emph{decomposition} of $G$ to be a partition of $V$ into sets
  $(A_{1}, \dotsc , A_{k})$, for some $k$.
We say that a decomposition is 
  a \emph{$\phi$-decomposition} if
  $\Conducin{A_{i}}{} \geq \phi$ for all $i$.
We define the boundary of a decomposition, written
  $\bdry{}{A_{1},\dotsc , A_{k}}$ to be the set of edges between different
  vertex sets in the partition:
\[
  \bdry{}{A_{1},\dotsc , A_{k}} = 
     E \intersect  \union_{i \not = j} (A_{i} \times A_{j}).
\]
We say that a decomposition $(A_{1}, \dotsc , A_{k})$ is a
  \emph{$\lambda$-spectral decomposition}
  if the smallest non-zero normalized Laplacian eigenvalue of
  $G (A_i)$ is at least $\lambda$, for all $i$.
By Cheeger's inequality (Theorem \ref{thm:cheeger}),
  every $\phi$-decomposition is a $(\phi^2/2)$-spectral decomposition.

\begin{theorem}\label{thm:graphDecompExist}
Let $G = (V,E)$ be a graph and let $m = \sizeof{E}$.
Then, $G$ has a $\left(6 \log_{4/3} 2 m \right)^{-1}$-decomposition with
$\sizeof{\bdry{}{A_{1}, \dotsc , A_{k}}} \leq \sizeof{E}/2$.
\end{theorem}

\subsection{Existence of spectral sparsifiers}\label{sec:existence}
Before proving Theorem \ref{thm:graphDecompExist},
  we first quickly explain how to use Theorem~\ref{thm:graphDecompExist}
  to prove that spectral sparsifiers exist.
Given any graph $G$, apply the theorem to find a decomposition of the graph
  into components of conductance $\Omega (1/\log n)$, with at most half of the
  original edges bridging components.
Because this decomposition is a $\Omega(1/\log^2 n)$-spectral decomposition,
  by Theorem~\ref{thm:sampling} we may sparsify the graph induced on
  each component by random sampling.
The average degree in the sparsifier for each component will be
  $O(\log^{6} n)$.
It remains to sparsify the edges bridging components.
If only $\softO{n}$ edges bridge components, then we do not need to sparsify
  them further.
If more edges bridge components, we sparsify them recursively.
That is, we treat those edges as a graph in their own right,
  decompose that graph, sample the edges induced in its components,
  and so on.
As each of these recursive steps reduces the number of edges remaining
  by at least a factor of two,
  at most a logarithmic number of recursive steps will be required, and thus the
  average degree of the sparsifier will be at most $O (\log^{7} n)$.
The above process also establishes the following decomposition theorem.


Recently, Batson, Spielman and Srivastava~\cite{BatsonSpielmanSrivastava}
  have shown that $(1+\epsilon )$-spectral sparsifiers with $O (n  / \epsilon^{2})$
  edges exist.

\subsection{The Proof of Theorem \ref{thm:graphDecompExist}}

Theorem \ref{thm:graphDecompExist} is not algorithmic.
It follows quickly from
  the following lemma, which says that if the largest set with conductance less
  than $\phi$ is small,
  then the graph induced on the complement has conductance almost $\phi$.
This lemma is the key component in our proof
  of Theorem~\ref{thm:graphDecompExist}, and its analog for 
  approximate sparsest cuts (Theorem \ref{thm:ApproxCut})
  is the key to our algorithm.

\begin{lemma}[Sparsest Cuts as Certificates]\label{lem:certificate}
Let $G = (V,E)$ be a graph and let $\phi \leq 1$.
Let $B \subseteq V$ and let $S \subset B$ be a set
  maximizing $\vol{S}$ among those
  satisfying
\begin{enumerate}
\item [(C.1)]  $\vol{S} \leq \vol{B}/2$, and
\item [(C.2)] $\conducin{B}{S} \leq \phi$.
\end{enumerate}
If $\vol{S} = \alpha  \vol{B}$ for $\alpha \leq 1/3$,
  then
\[
  \Conducin{B-S}{} \geq \phi \left(\frac{1-3 \alpha}{1 - \alpha } \right).
\]
\end{lemma}
\begin{proof}
Let $S$ be a set of maximum size that satisfies $(C.1)$ and $(C.2)$,
  let
\[
\beta = \frac{1-3\alpha}{1-\alpha},
\]
and
  assume by way of contradiction that
  $\Conducin{B-S}{} < \phi \beta .$
Then, there exists a set
  $R \subset B-S$
  such that
\[
  \conducin{B-S}{R} < \phi \beta , \text{ and}
\]
\[
  \vol{R} \leq \frac{1}{2}\vol{B-S}.
\]
Let $T = R \union S$.
We will prove
\[
  \conducin{B}{T} < \phi
\]
and $\vol{S} \leq \min \left(\vol{T}, \vol{B-T} \right)$,
  contradicting the maximality of $S$.

We begin by observing that
\begin{align}
  \sizeof{E (T, B - T) } = 
  \sizeof{E (R \union S, B - (R \union S))}
& \leq
 \sizeof{E (S, B - S)}
+
 \sizeof{E (R, B - S - R))} \notag \\
& <
 \phi \vol{S}
+
 (\phi \beta ) \vol{R}. \label{eqn:certificate1}
\end{align}

We divide the rest of our proof into two cases, depending on whether
  or not $\vol{T} \leq \vol{B}/2$.
First, consider the case in which
  $\vol{T} \leq \vol{B}/2$.
In this case, $T$ provides a contradiction to the maximality of
  $S$, as $\vol{S} < \vol{T} \leq \vol{B}/2$,
  and
\[
  \sizeof{E (T, B - T) }
< \phi \left(\vol{S} + \vol{R} \right) = \phi \vol{T},
\]
which implies
\[
  \conducin{B}{T} < \phi .
\]

In the case $\vol{T} > \vol{B}/2$,
  we will prove that the set $B - T$ contradicts the maximality of $S$.
First, we show
\begin{equation}\label{eqn:certificate2}
\vol{B-T} > \left( \frac{1-\alpha}{2} \right) \vol{B},
\end{equation}
which implies
  $\vol{B-T} > \vol{S}$ because we assume $\alpha \leq 1/3$.
To prove \eqref{eqn:certificate2}, compute
\begin{align*}
  \vol{T} & = \vol{S} + \vol{R}\\
 &  \leq \vol{S} + (1/2) (\vol{B} - \vol{S})\\
 & = (1/2) \vol{B} + (1/2) \vol{S}\\
 & = \left(\frac{1+\alpha}{2} \right) \vol{B}.
\end{align*}

To upper bound the conductance of $T$, we compute
\begin{align*}
  \sizeof{E (T, B-T)}
& <
 \phi \vol{S}
+
 (\phi \beta) \vol{R} \quad \text{(by \eqref{eqn:certificate1})}
\\
& \leq
 \phi \vol{S}
+
 (\phi \beta ) (\vol{B} - \vol{S}) / 2
\\
& =
\phi \vol{B} \left(
  \alpha + \beta (1 - \alpha)/2
 \right).
\end{align*}

So,
\[
\conducin{B}{T}
=
\frac{
  \sizeof{E (T, B-T)}
 }{
   \min (\vol{T}, \vol{B-T})
 }
=
\frac{
  \sizeof{E (T, B-T)}
 }{
   \vol{B-T}
 }
\leq
\frac{
  \phi \vol{B}  \left( \alpha + \beta (1 - \alpha)/2 \right)
 }{
   \vol{B} (1-\alpha)/2
 }
=
\phi,
\]
by our choice of $\beta$.
\end{proof}

We will prove Theorem~\ref{thm:graphDecompExist} by proving that
  the following procedure produces the required decomposition.

\vskip 0.2in
\noindent
\fbox{
\begin{minipage}{6in}
Set $\phi = \left(2 \log_{4/3} \vol{V} \right)^{-1}$.\\
Note that we initially call this algorithm with $B = V$.
\\

\noindent $\mathtt{idealDecomp} (B, \phi )$
\begin{enumerate}

\item [1.] If $\Conducin{B}{} \geq \phi$, then return $B$.  Otherwise, proceed.

\item [2.] Let $S$ be the subset of $B$ maximizing $\vol{S}$ satisfying (C.1) and (C.2).

\item [3.] If $\vol{S} \leq \vol{B}/4$, return the decomposition
   $(B-S, \texttt{idealDecomp} (S , \phi ))$,

\item [4.] else, return the decomposition
  $(\texttt{idealDecomp} (B - S , \phi ), \texttt{idealDecomp} (S , \phi ))$.
\end{enumerate}
\end{minipage}
}
\vskip 0.2in

\begin{proof}[Proof of Theorem~\ref{thm:graphDecompExist}]
To see that the recursive procedure terminates, recall that we have defined
  $\Conducin{B}{} = 1$ when $\sizeof{B} = 1$.

Let $(A_{1}, \dotsc , A_{k})$ be the output of $\mathtt{idealDecomp} (V)$.
Lemma~\ref{lem:certificate} implies that
  $\Conducin{A_{i}}{} \geq \phi/3$ for each $i$.

To bound the number of edges in $\bdry{}{A_{1}, \dotsc , A_{k}}$,
  note that the depth of the recursion is at most $\log_{4/3} \vol{V}$
  and that at most a $\phi$ fraction of the edges are added to
  $\bdry{}{A_{1}, \dotsc , A_{k}}$ at each level of the recursion.
So, 
\[
\sizeof{\bdry{}{A_{1}, \dotsc , A_{k}}} \leq \sizeof{E} \phi \log_{4/3} \vol{V}
 \leq \sizeof{E}/2.
\]
\end{proof}

\section{Approximate Sparsest Cuts}\label{sec:approxCut}

Unfortunately, it is NP-hard to compute sparsest cuts.
So, we cannot directly
  apply Lemma~\ref{lem:certificate} in the design of our algorithm.
Instead, we will apply a nearly-linear time algorithm, \approxcut,
  that computes approximate sparsest cuts that satisfy an analog of
  Lemma~\ref{lem:certificate}, stated in Theorem~\ref{thm:ApproxCut}.
Whereas in Lemma~\ref{lem:certificate} we proved that if the largest sparse cut
  is small then its
  complement has high conductance,
  here we prove that if the cut output by \approxcut\ is small, then
  its complement is contained in a subgraph of high conductance.

The algorithm \approxcut \ works by repeatedly calling a routine
  for approximating sparsest cuts, \partition, from~\cite{SpielmanTengCuts}.
On input a graph that contains a sparse cut, with high probability
  the algorithm \partition \ either finds a large cut or a cut that has high overlap
  with the sparse cut.
We have not been able to find a way to quickly
   use an algorithm satisfying such a guarantee
  to certify that the complement of a small cut has high conductance.
Kannan, Vempala and Vetta~\cite{KannanVempalaVetta} showed that if we applied
  such an algorithm until it could not find any more cuts then we
  could obtain such a guarantee.
However, such a procedure could require quadratic time, which it too
  slow for our purposes.

\begin{theorem}[\approxcut]\label{thm:ApproxCut}
Let $\phi, p \in (0,1)$ and let $G = (V,E)$ be a graph with $m$ edges.
Let $D$ be the output of $\approxcut (G, \phi , p)$.
Then
\begin{itemize}
\item [(A.1)] $\vol{D} \leq (23/25) \vol{V}$,
\item [(A.2)] If $D \not = \emptyset$ then $\conduc{G}{D} \leq  \phi $, and
\item [(A.3)] With probability at least $1-p$, either
\begin{itemize}
\item [(A.3.a)] $\vol{D} \geq (1/29) \vol{V}$, or
\item [(A.3.b)] there exists a set $W \supseteq V-D$ for which
  $\Conducin{W} \geq f_{2} (\phi)$, where
\begin{equation}\label{eqn:f4}
  f_{2} (\phi) \defeq \frac{c_{2} \phi^{2}}{\log^{4} m},
\end{equation}
for some absolute constant $c_{2}$.
\end{itemize}
\end{itemize}
Moreover, the expected running time of \approxcut\  is 
  $O \left(\phi^{-4}m \log^{9} m \log (1/p) \right) $.
\end{theorem}

The code for \approxcut\ follows.
It relies on a routine called \partitiontwo\, which in turn relies on a routine called
  \partition\ from~\cite{SpielmanTengCuts}.
While one could easily combine the routines \approxcut \ and \partitiontwo ,
  their separation simplifies our analysis.
The algorithm \partitiontwo \ is very simple: it just calls 
  \partition \ repeatedly and collects the cuts it produces until they
  contain at least $1/5$ of the volume of the graph or until it has
  made enough calls.
The algorithm \approxcut \ is similar: it calls \partitiontwo \ in the
  same way that \partitiontwo \ calls \partition .
  
\vskip 0.2in
\noindent
\fbox{
\begin{minipage}{6in}
\noindent $D =\approxcut(G, \phi , p)$,
  where $G$ is a graph, $\phi, p,  \in (0,1)$.
\begin{enumerate}
\item [(0)] Set $V_{0} = V$ and $j = 0$.

\item [(1)] Set $r = \ceiling{\log_{2} (m)}$ and $\epsilon = \min (1/2r, 1/5)$.
\item [(2)] While $j <  r$ and
$\vol{V_{j}} \geq (4/5) \vol{V}$,
\begin{enumerate}
\item [(a)] Set $j = j + 1$.
\item [(b)] Set $D_{j} = \partitiontwo (G \{V_{j-1} \}, (2/23)\phi , p/2r, \epsilon )$
\item [(c)] Set $V_{j} = V_{j-1} - D_{j}$.
\end{enumerate}
\item [(3)] Set $D = D_{1} \union \dotsb \union D_{j}$.
\end{enumerate}
\end{minipage}
}
\vskip 0.2in

\subsection{Partitioning in Nearly-Linear-Time}

\vskip 0.2in
\noindent
\fbox{
\begin{minipage}{6in}
\noindent $D =\partitiontwo(G, \theta , p, \epsilon )$,
where $G$ is a graph, $\theta, p,  \in (0,1)$ and $\epsilon \in (0,1)$.
\begin{enumerate}
\item [(0)] Set $W_{0} = V$ and $j = 0$.  Set $r = \ceiling{\log_{2} (1/\epsilon )}$.

\item [(1)] While $j <  r$ and
$\vol{W_{j}} \geq (4/5) \vol{V}$,
\begin{enumerate}
\item [(a)] Set $j = j + 1$.
\item [(b)] Set $D_{j} = \partition (G \{W_{j-1} \}, \theta/9 , p/r)$
\item [(c)] Set $W_{j} = W_{j-1} - D_{j}$.
\end{enumerate}
\item [(2)] Set $D = D_{1} \union \dotsb \union D_{j}$.
\end{enumerate}
\end{minipage}
}
\vskip 0.2in

The algorithm \partition\ from~\cite{SpielmanTengCuts},
  satisfies the following theorem (see \cite[Theorem 3.2]{SpielmanTengCuts})

\begin{theorem}[\partition]\label{thm:Partition}
Let $D$ be the output of $\partition (G, \tau , p)$,
  where $G$ is a graph and $\tau, p \in (0,1)$.
Then
\begin{itemize}
\item [(P.1)] $\vol{D} \leq (7/8) \vol{V}$,
\item [(P.2)] If $D \not = \emptyset$ then $\conduc{G}{D} \leq  \tau $, and
\item [(P.3)] For some absolute constant $c_{1}$ and
\[
  f_{1} (\tau) \defeq  \frac{c_{1} \tau^{2}}{\log^{3} m},
\]
for \textbf{every} set $S$ satisfying
\begin{equation}\label{eqn:P3}
\vol{S} \leq \vol{V}/2 \quad \text{and} \quad  \conduc{G}{S} \leq f_{1} (\tau),
\end{equation}
with probability at least $1-p$ either
\begin{itemize}
\item [(P.3.a)] $\vol{D} \geq (1/4) \vol{V}$, or
\item [(P.3.b)] $\vol{S \intersect D} \geq \vol{S}/2$.
\end{itemize}
\end{itemize}
Moreover, the expected running time of \partition \ is 
  $O \left(\tau^{-4}m \log^{7} m \log (1/p) \right) $.
\end{theorem}
If either $(P.3.a)$ or $(P.3.b)$ occur for a set
  $S$ satisfying \eqref{eqn:P3}, we say that \partition \
 \textbf{succeeds} for $S$.
Otherwise, we say that it \textbf{fails}.

One can view condition $(A.3)$ in Theorem~\ref{thm:ApproxCut} as reversing
  the quantifiers in condition $(P.3)$ in Theorem~\ref{thm:Partition}.
Theorem~\ref{thm:Partition} says that for every set $S$ of low conductance
  there is a good probability that a substantial portion of $S$ is removed.
On the other hand, Theorem~\ref{thm:ApproxCut} says that with high probability
  all sets of low conductance will be removed.

The algorithm \partitiontwo \ satisfies a guarantee similar to that of \partition,
  but it strengthens condition $(P.3.b)$.

\begin{lemma}[\partitiontwo]\label{lem:partition2}
Let $D$ be the output of $\text{\partitiontwo} (G, \theta , p, \epsilon )$,
  where $G$ is a graph, $\theta, p \in (0,1)$ and $\epsilon \in (0,1)$.
Then
\begin{itemize}
\item [(Q.1)] $\vol{D} \leq (9/10) \vol{V}$,
\item [(Q.2)] If $D \not = \emptyset$ then $\conduc{G}{D} \leq \theta $, and
\item [(Q.3)] For \textbf{every} set $S$
  satisfying
\begin{equation}\label{eqn:Q3}
\vol{S} \leq \vol{V}/2 \quad \text{and} \quad 
  \conduc{G}{S} \leq f_{1} (\theta/9),
\end{equation}
  with probability at least $1-p$,
  either
\begin{itemize}
\item [(Q.3.a)] $\vol{D} \geq (1/5) \vol{V}$, or
\item [(Q.3.b)] $\vol{S \intersect D} \geq (1-\delta  )\vol{S}$,
  where $\delta = \max \left(\epsilon ,   \conduc{G}{S}/ f_{1} (\theta/9) \right)$.
\end{itemize}
\end{itemize}
Moreover, the expected running time of \partitiontwo \ is
  $O \left(\theta^{-4} m \log^{7} m \log (1/\epsilon )\log (\log (1/\epsilon ) /p) \right) $.
\end{lemma}
If either $(Q.3.a)$ or $(Q.3.b)$ occur
  for a set $S$ satisfying \eqref{eqn:Q3},
  we say that \partitiontwo \
 \textbf{succeeds} for $S$.
Otherwise, we say that it \textbf{fails}.

The proof of this lemma is routine, given Theorem~\ref{thm:Partition}.

\begin{proof}
Let $j^{*}$ be such that $D = D_{1} \union \dotsb \union D_{j^{*}}$.
To prove (Q.1),
  let $\nu = \vol{(D_{1} \union \dotsb \union D_{j^{*}-1})}/\vol{V}$.
As $\vol{W_{j^{*}-1}} \geq (4/5) \vol{V}$, $\nu \leq 1/5$.
By $(P.1)$,
  $\vol{D_{j^{*}}} \leq  (7/8) \vol{W_{j^{*}-1}}$, so
\[
\vol{D_{1} \union \dotsb \union D_{j^{*}}}
\leq \vol{V} (\nu + (7/8) (1-\nu))
\leq \vol{V} ((1/5) + (7/8) (4/5))
  = (9/10) \vol{V}.
\]

To establish (Q.2),
  we first compute
\begin{align*}
\sizeof{E (D, V-D)}
& =
\sum_{i=1}^{j^{*}} \sizeof{E (D_{i}, V-D)}\\
& \leq
\sum_{i=1}^{j^{*}}
  \sizeof{E (D_{i}, W_{i-1} - D_{i})}\\
& \leq
\sum_{i=1}^{j^{*}}
  (\theta /9) \min \left(\vol{D_{i}}, \vol{W_{i-1} - D_{i}} \right)
\quad
\text{(by (P.2) and line 1b of \partitiontwo  )}\\
& \leq
\sum_{i=1}^{j^{*}}
  (\theta/9)  \vol{D_{i}}
\\
& = (\theta/9)  \vol{D}.
\end{align*}
So, if  $\vol{D} \leq \vol{V}/2$,
  then $ \conduc{G}{D} \leq \theta/9$.
On the other hand, we established above that
  $\vol{D} \leq (9/10) \vol{V}$, from which
  it follows that
\[
  \vol{V-D} \geq (1/10) \vol{V} \geq (1/10) (10/9) \vol{D} = (1/9) \vol{D}.
\]
So,
\[
  \conduc{G}{D}
=
 \frac{\sizeof{E (D, V-D)}}
      {\min \left(\vol{D}, \vol{V-D} \right)}
\leq
9  \frac{\sizeof{E (D, V-D)}}{\vol{D}}
\leq
 \theta.
\]

To prove (Q.3),
  let $S$ be a set satisfying \eqref{eqn:Q3}, and
  let $S_{j} = S \intersect W_{j}$.
From Theorem~\ref{thm:Partition}, we know that with probability at least
  $1 - p/r$,
\begin{equation}\label{eqn:part2half}
 \vol{S_{1}} \leq (1/2) \vol{S_{0}}.
\end{equation}
We need to prove that with probability at least $1-p$,
  either $\vol{W_{j^{*}}} \leq (4/5) \vol{V}$
  or
  $\vol{S_{j^{*}}} \leq \delta \vol{S}$.
If neither of these inequalities hold, then
\[
  j^{*} = r,
\quad \vol{W_{r}} \geq (4/5) \vol{V},
\quad \text{and} \quad \vol{S_{r}} > \delta \vol{S} \geq  \epsilon \vol{S},
\]
where we recall $r = \ceiling{\log_{2} (1/\epsilon )}$.
So, there must exist a $j$ for which $\vol{S_{j+1}} \geq (1/2) \vol{S_{j}}$.
If $S_{j}$ satisfied condition \eqref{eqn:Q3} 
  in $G\{V_{j} \}$ this would imply that
  \partition \ failed for $S_{j}$.
We already know this is unlikely for $j=0$.
To show it is unlikely for $j \geq 1$, we prove that $S_{j}$ does
  satisfy condition \eqref{eqn:Q3} 
  in $G\{V_{j}\}$.
Assuming \eqref{eqn:part2half},
\begin{multline*}
\conduc{G\{W_{j} \}}{S_{j}}
=
\conducin{W_{j}}{S_{j}}
=
\frac{
  \sizeof{\bdry{W_{j}}{S_{j}}}
}{
  \min \left(\vol{S_{j}}, \vol{W_{j} - S_{j}} \right)
}
=
\frac{
  \sizeof{\bdry{W_{j}}{S_{j}}}
}{
  \vol{S_{j}}
}
\leq
\frac{
  \sizeof{\bdry{V}{S}}
}{
  \vol{S_{r}}
}\\
\leq
\frac{
  \sizeof{\bdry{V}{S}}
}{
  \delta \vol{S}
}
=
(1/\delta)\conduc{G}{S}
\leq
f_{1} (\theta/9),
\end{multline*}
where the third equality follows from the assumption 
  $\vol{S_{1}} \leq (1/2) \vol{S_{0}} \leq (1/4) \vol{V}$ and the last
  inequality follows from the definition
  $\delta = \max \left(\epsilon ,   \conduc{G}{S}/ f_{1} (\theta/9) \right)$.
So, $S_{j}$ satisfies conditions \eqref{eqn:P3} 
  with $\tau = \theta/9$, but \partition \ fails for $S_{j}$.
As there are at most $r$ sets $S_{j}$, this happens for one of them with probability at most
  $r (p/r) = p$.

Finally, the bound on the expected running time of \partitiontwo \
  is immediate from the bound on the running time of \partition.
\end{proof}

\subsection{Proof of Theorem~\ref{thm:ApproxCut}}
The rest of this section is devoted to the proof of Theorem~\ref{thm:ApproxCut}, with all but one
  line devoted to part (A.3).
Our goal is to prove the existence of a set of vertices $W$ of high conductance
  that contains all the vertices not cut out by \approxcut .
We will construct this set $W$ in stages.
Recall that $V_{i} = V - D_{1} \union \dotsb \union D_{i}$ is the set of vertices
  that are not removed by the first $i$ cuts.
In stage $i$, we will express $W_{i}$, a superset of $V_{i}$, as a set of high
  conductance $U_{i-1}$ plus some vertices in $V_{i}$.
We will show that in each stage the volume of the vertices that are
  not in the set of high conductance shrinks by at least a factor of 2.

We do this by letting $S_{i}$ be the biggest set of conductance at most
  $\sigma_{i}$ in $W_{i}$, where $\sigma_{i}$ is a factor $(1-2 \epsilon)$ smaller
  than the conductance of $U_{i-1}$.
We then show that at least a $2\epsilon$ fraction of the volume
  of $S_{i}$ lies outside $U_{i-1}$ and thus inside $V_{i}$.
From Lemma~\ref{lem:certificate} we know that $U_{i} \defeq  W_{i} - S_{i}$
  has high conductance.
We will use Lemma~\ref{lem:partition2} to show that at most an $\epsilon$
  fraction of $S_{i}$ appears in $V_{i+1}$.
So, the volume of $S_{i}$ that remains inside $V_{i+1}$ will be at most
  half the volume of $V_{i}$ that is not in $U_{i-1}$.
We then set $W_{i+1} = U_{i} \union (S_{i} \intersect V_{i+1})$, and proceed
  with our induction.
Eventually, we will arrive at an $i$ for which either $W_{i}$ 
  has high conductance or enough volume has been removed from $V_{i}$.

\begin{figure}[h]
\centering
\subfigure[The subsets of $W_{i}$.
Not drawn to scale.]{\epsfig{figure=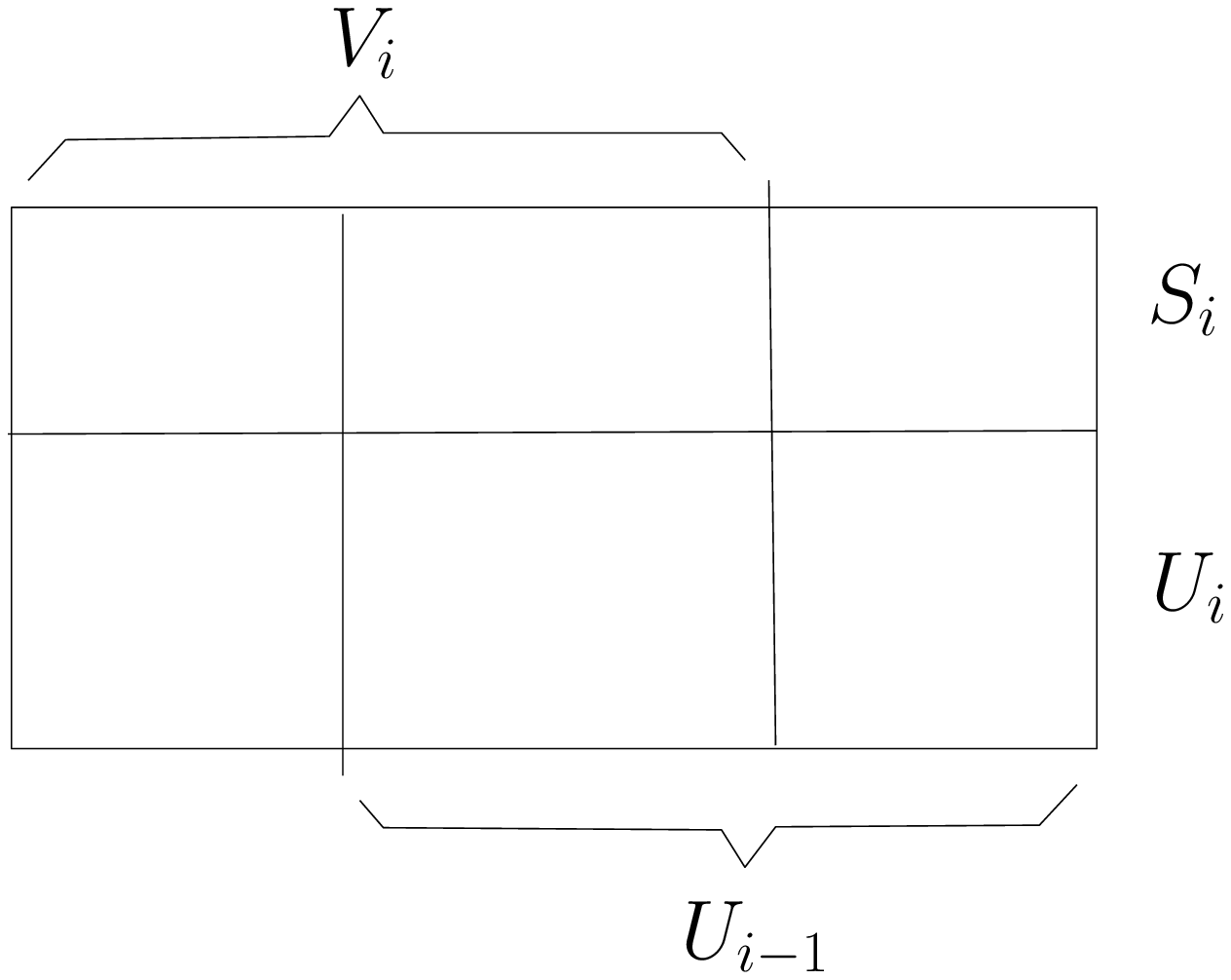,width=2in}} \qquad
\subfigure[The shaded portion is $W_{i+1}$.
It equals $U_{i} \union V_{i+1}$, and so can be viewed as the
union of the set of vertices maintained by the algorithm with the
high-conductance set we know exists.]{\epsfig{figure=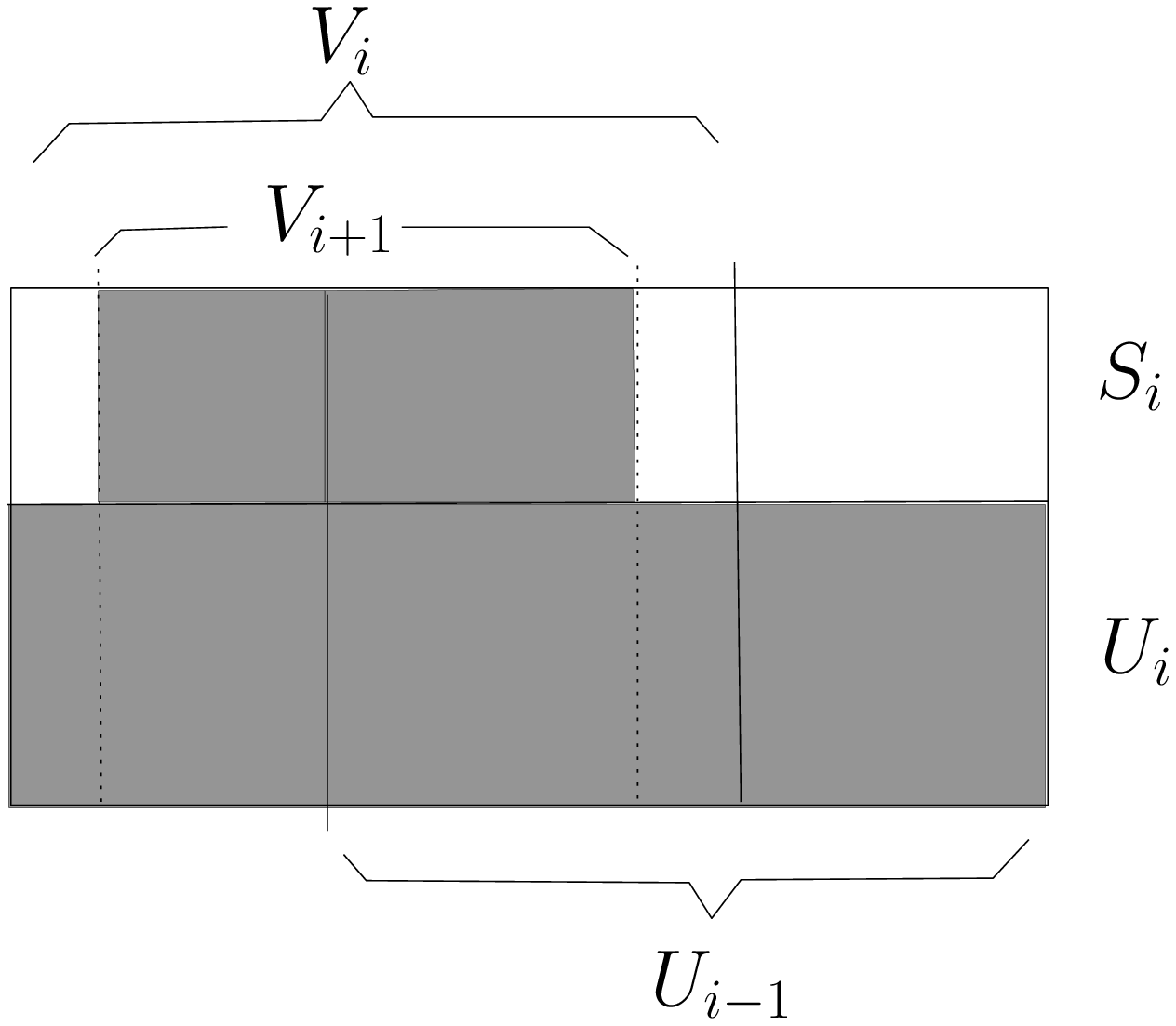,width=2in}}
\end{figure}

Formally, we set 
\[
W_{0} = V_{0} = V
\quad
\text{and}
\quad 
\sigma_{0} = \epsilon  f_{1} (\phi / 104).
\]
We then construct sets $S_{i}$, $U_{i}$ and $W_{i}$ by the following
  inductive procedure.

\begin{itemize}

\item [1.] Set $i = 0$.

\item [2.] While $i \leq r$ and $W_{i}$ is defined,

\begin{itemize}

\item [a.] If $W_{i}$ contains a set $S_{i}$
 such that
\[
  \vol{S_{i}} \leq (1/2) \vol{W_{i}}
\quad
\text{and}
\quad
\conducin{W_{i}}{S_{i}} \leq \sigma_{i},
\]
set $S_{i}$ to be such a set of maximum size.

If $\vol{S_{i}} \geq (2/17) \vol{V}$, 
  stop the procedure and leave $W_{i+1}$ undefined.

If there is no such set, set $S_{i} = \emptyset$, set $U_{i} = W_{i}$,
  stop the procedure and leave $W_{i+1}$ undefined.

\item [b.] Set $U_{i} = W_{i} - S_{i}$.
\item [c.] Set $\theta_{i} =
   \left(1 - 3 \frac{\vol{S_{i}}}{\vol{W_{i}}} \right) \sigma_{i}$.

\item [d.] Set $\sigma_{i+1} = (1 - 2 \epsilon) \theta_{i}$.

\item [e.] Set $W_{i+1} = U_{i} \union (S_{i} \intersect  V_{i+1})$.

\item [f.] Set $i = i + 1$.


\end{itemize}

\item [3.] Set $W = W_{i}$ where $i$ is the last index for which $W_{i}$ is defined.
\end{itemize}

Note that there may be many choices for a set $S_{i}$.
Once a choice is made, it must be fixed for the rest of the procedure so that
  we can reason about it using Lemma~\ref{lem:partition2}.

We will prove that if some set $S_{i}$ has volume
  greater than $(2/17) \vol{V}$, then with high probability \approxcut \
  will return a large cut $D$, and hence part (A.3.a) is satisfied.
Thus, we will be mainly concerned with the case in which this does not happen.
In this case, we will prove that $\theta_{i}$ is not too much less than $\sigma_{0}$,
and so the set $U_{i}$ has high conductance.
If the procedure stops because $S_{i}$ is empty, then $W_{i} = U_{i}$
  is the set of high conductance we seek.
We will prove that for some $i \leq r$ probably either $S_{i}$ is empty, 
  $\vol{S_{i}} \geq (2/17) \vol{V}$ or $\vol{V_{i}} \leq (16/17) \vol{V}$.

\begin{claim}\label{clm:VinsideW}
For all $i$ such that $W_{i+1}$ is defined,
\[
  V_{i+1} \subseteq W_{i+1} \subseteq W_{i}.
\]
\end{claim}
\begin{proof}
We prove this by induction on $i$.
For $i = 0$, we know that $V_{i} = W_{i}$.
As $W_{i} = U_{i} \union S_{i}$ and the algorithm ensures $V_{i+1} \subseteq V_{i}$,
\[
  V_{i+1} \subseteq V_{i} \subseteq W_{i} = U_{i} \union S_{i}.
\]
Thus,
\[
  V_{i+1} \subseteq U_{i} \union (S_{i} \intersect V_{i+1})
  = W_{i+1}
  \subseteq U_{i} \union S_{i}
  = W_{i}.
\]
\end{proof}

\begin{claim}\label{clm:uiExpander}
For all $i$ such that $U_{i}$ is defined
\[
\Conducin{U_{i}} \geq \theta_{i}.
\]
\end{claim}
\begin{proof}
Follows immediately from  Lemma~\ref{lem:certificate} and
  the definitions of $S_{i}$ and $\theta_{i}$.
\end{proof}

\begin{lemma}\label{lem:ac1}
If
\begin{itemize}
\item [(a)] $\vol{S_{i}} \leq (2/17) \vol{V}$, and
\item [(b)] $\vol{V_{i-1}} \geq (16/17) \vol{V}$, then
\end{itemize}
  then
\[
\vol{S_{i} \intersect (S_{i-1} \intersect V_{i})} \geq 2 \epsilon \vol{S_{i}}.
\]
\end{lemma}
\begin{proof}
This lemma follows easily from the definitions of the sets 
  $S_{i}$, $U_{i}$ and $V_{i}$.
As $V_{i-1} \subseteq W_{i-1}$ and $\vol{U_{i-1}} \geq (1/2) \vol{W_{i-1}}$,
\[
  \vol{U_{i-1}} \geq (1/2) \vol{V_{i-1}}
  \geq (8/17) \vol{V}
 \geq 4 \vol{S_{i}}.
\]
So, we may apply Claim~\ref{clm:uiExpander} 
  to show
\[
 \sizeof{ \bdry{U_{i-1}}{S_{i}}}
  \geq 
 \sizeof{ \bdry{U_{i-1}}{S_{i} \intersect U_{i-1}}}
   \geq \theta_{i-1} \vol{S_{i} \intersect U_{i-1}}.
\]
On the other hand, 
\[
 \sizeof{  \bdry{U_{i-1}}{S_{i}}}
\leq 
 \sizeof{  \bdry{W_{i}}{S_{i}}}
\leq 
  \sigma_{i} \vol{S_{i}}  
 =
  (1-2\epsilon) \theta_{i-1} \vol{S_{i}}.
\]
Combining these two inequalities yields
\[
\theta_{i-1} \vol{S_{i} \intersect U_{i-1}}
\leq 
  (1-2\epsilon) \theta_{i-1} \vol{S_{i}}
\]
and
\[
 \vol{S_{i} \intersect U_{i-1}}
\leq 
  (1-2\epsilon) \vol{S_{i}}.
\]
As
\[
  S_{i} \subseteq W_{i} = U_{i-1} \union (S_{i-1} \intersect V_{i}),
\]
we may conclude
\[
  \vol{S_{i} \intersect (S_{i-1} \intersect V_{i})} 
\geq
  2 \epsilon \vol{S_{i}}.
\]
\end{proof}

We now show that if at most an $\epsilon$ fraction of each $S_{i}$ appears in
  $V_{i+1}$, then the sets $S_{i} \intersect V_{i+1}$ shrink to the point of vanishing.

\begin{lemma}\label{lem:ac2}
If all defined $S_{i}$ and $V_{i}$ satisfy
\begin{itemize}
\item [(a)] $\vol{S_{i}} \leq (2/17) \vol{V}$, 
\item [(b)] $\vol{V_{i}} \geq (16/17) \vol{V}$, and
\item [(c)] $\vol{S_{i} \intersect V_{i+1}} \leq \epsilon \vol{S_{i}}$,
\end{itemize}
then
for all $i \geq 1$ for which $S_{i}$ is defined,
\[
  \vol{S_{i} \intersect V_{i+1}} \leq (1/2) \vol{S_{i-1} \intersect V_{i}},
\]
and
\[
  \vol{S_{i}} \leq (1/2) \vol{S_{i-1}}.
\]
In particular, the set $S_{r}$ is empty if it is defined.
\end{lemma}
\begin{proof}
Lemma~\ref{lem:ac1} tells us that
\[
  \epsilon \vol{S_{i}} 
  \leq (1/2) \vol{S_{i} \intersect (S_{i-1} \intersect V_{i})}
  \leq (1/2) \vol{ S_{i-1} \intersect V_{i}}.
\]
Combining this inequality with $(c)$ yields
\[
  \vol{S_{i} \intersect V_{i+1}} \leq (1/2) \vol{S_{i-1} \intersect V_{i}}.
\]
Similarly, we may conclude from Lemma~\ref{lem:ac1} that
\[
  \epsilon \vol{S_{i+1}} \leq (1/2) \vol{S_{i} \intersect V_{i+1}},
\]
which when combined with $(c)$ yields
\[
  \epsilon \vol{S_{i+1}} \leq (1/2) \epsilon \vol{S_{i}},
\]
from which the second part of the lemma follows.

For $S_{i}$ to be defined, we must have $\vol{S_{0}} \leq (2/17) \vol{V}$;
  so,
\[
\vol{S_{r}} \leq (1/2)^{r} \vol{S_{0}} 
  \leq (1/2)^{\ceiling{\log_{2} \vol{V}/2 }} (2/17) \vol{V}
  \leq \frac{2}{\vol{V}} (2/17) \vol{V} < 1.
\]
We conclude that the set $S_{r}$ must be empty if it is defined.
\end{proof}

This geometric shrinking of the volumes of the sets $S_{i}$ allows
  us to prove a lower bound on $\theta_{i}$.

\begin{lemma}\label{lem:ac3}
Under the conditions of Lemma~\ref{lem:ac2},
\[
  \theta_{i} \geq 
\frac{c_{2} \phi^{2}}{\log^{4} m},
\]
for some absolute constant $c_{2}$.
\end{lemma}
\begin{proof}
We have
\begin{align*}
  \theta_{i}
  & = \sigma_{0} (1-2 \epsilon)^{i-1} 
  \prod_{j=0}^{i} \left(1 - \frac{3 \vol{S_{j}}}{\vol{W_{j}}} \right).
\end{align*}
As $i \leq r$ and $\epsilon = \min (1/5, 1/2r)$, we have
\[
  (1-2 \epsilon)^{i-1} \geq 1/e.
\]
To analyze the other product, we apply Lemma~\ref{lem:ac2} to prove
\[
  \sum_{j=0}^{i} \vol{S_{j}} \leq 2 \vol{S_{0}},
\]
and so
\begin{align*}
\prod_{j=0}^{i} \left(1 - \frac{3 \vol{S_{j}}}{\vol{W_{j}}} \right)
& \geq
1 -   \sum_{i = 0}^{r} \frac{3 \vol{S_{i}}}{(16/17)\vol{V}}
\\
& \geq
1 -  \frac{2 \cdot  3 \cdot  17}{16} \frac{\vol{S_{0}}}{\vol{V}},
& 
\\
& \geq
1 -  \frac{2 \cdot  3 \cdot  17}{16} \frac{2}{17}
& 
\\
& = \frac{1}{4}.
\end{align*}
Thus,
\[
\theta_{i} 
\geq 
\frac{\sigma_{0}}{4 e}
\geq 
\frac{\epsilon f_{1} (\phi /104)}{4 e}
\geq 
\frac{c_{1} \phi^{2}}{4 e (104)^{2} \ceiling{\log m} \log^{3} m}
\geq 
\frac{c_{2} \phi^{2}}{\log^{4} m},
\]
for some constant $c_{2}$.
\end{proof}

To prove that condition $(c)$ of Lemma~\ref{lem:ac2} is probably satisfied,
  we will consider two cases.
First, if $\vol{S_{i} \intersect V_{i}} \leq \epsilon \vol{S_{i}}$
  then $(c)$ is trivially satisfied as $V_{i+1} \subseteq V_{i}$.
On the other hand, if $\vol{S_{i} \intersect V_{i}} \geq \epsilon \vol{S_{i}}$,
  then we will show that $S_{i} \intersect V_{i}$ satisfies conditions \eqref{eqn:Q3}
  in $G \{V_{i} \}$, and so 
  with high probability the
  cut $D_{i+1}$ made by 
  \partitiontwo \ removes enough of $S_{i}$.

\begin{lemma}\label{lem:ac4}
If
\begin{itemize}
\item [(a)] $\vol{S_{i}} \leq (2/17) \vol{V}$, 
\item [(b)] $\vol{V_{i}} \geq (16/17) \vol{V}$, and
\item [(c)] $\vol{S_{i} \intersect V_{i}} \geq \epsilon \vol{S_{i}}$,
\end{itemize}
then
\[
  \conduc{G \{V_{i} \} }{S_{i} \intersect V_{i}} 
  \leq \frac{\epsilon}{\delta} f_{1} (\phi/104),
\]
where $\delta = \vol{S_{i} \intersect V_{i}} / \vol{S_{i}}$.
If, in addition 
\[
\vol{S_{i} \intersect V_{i+1}} 
  \leq \frac{\epsilon}{\delta } \vol{S_{i} \intersect V_{i}},
\]
then
\[
\vol{S_{i} \intersect V_{i+1}} \leq \epsilon \vol{S_{i}}.
\]
\end{lemma}
\begin{proof}
By Claim~\ref{clm:ac0},
\[
  \sizeof{\bdry{V_{i}}{S_{i} \intersect V_{i}}}
\leq 
  \sizeof{\bdry{W_{i}}{S_{i}}}.
\]
Set $\delta  = \vol{S_{i} \intersect V_{i}} / \vol{S_{i}}$.
Assumption $(c)$ tells us that $\delta \geq \epsilon$.
As $\vol{S_{i}} \leq (1/2) \vol{V_{i}}$, 
\[
  \conduc{G \{V_{i} \}}{S_{i} \intersect V_{i}}
=
\frac{
  \sizeof{\bdry{V_{i}}{S_{i} \intersect V_{i}}}
}{
  \vol{S_{i} \intersect V_{i}}
}
\leq 
\frac{
  \sizeof{ \bdry{W_{i}}{S_{i}} }
}{
  \delta \vol{S_{i}}
}
=
\frac{1}{\delta}
\conduc{G \{W_{i} \}}{S_{i}}
\leq 
  \frac{\sigma_{i}}{\delta}
=
  \frac{\sigma_{i}}{\epsilon}
  \frac{\epsilon}{ \delta }
\leq 
  \frac{\sigma_{0}}{\epsilon}
  \frac{\epsilon}{ \delta }
=
  \frac{\epsilon}{\delta} f_{1} (\phi /104).
\]
The last part of the lemma is trivial.
\end{proof}

\begin{claim}\label{clm:ac0}
\[
  \bdry{V_{i}}{S_{i} \intersect V_{i}}
\subseteq 
  \bdry{W_{i}}{S_{i}}.
\]
\end{claim}
\begin{proof}
\[
  \bdry{V_{i}}{S_{i} \intersect V_{i}}
=
 E (S_{i} \intersect V_{i}, V_{i} - (S_{i} \intersect V_{i}))
\subseteq 
 E (S_{i} , V_{i} - (S_{i} \intersect V_{i}))
\subseteq 
 E (S_{i} , W_{i} - (S_{i} \intersect W_{i}))
=
  \bdry{W_{i}}{S_{i}}.
\]
\end{proof}

We now show that if $\vol{S_{i}} \geq (2/17) \vol{V}$,
  then in the $i$th iteration \partition2 \ will probably
  remove a large portion of the graph.
If $\vol{S_{i} \intersect V_{i}} \leq (1/2) \vol{V_{i}}$
  we will argue that $S_{i} \intersect V_{i}$ satisfies condition
  \eqref{eqn:Q3} in $G \{V_{i}  \}$.
Otherwise, will argue that $V_{i} - S_{i} \intersect V_{i}$ does.

\begin{lemma}\label{lem:ac5}
If
\begin{itemize}
\item [(a)] $\vol{V_{i}} \geq (16/17) \vol{V}$,
\item [(b)] $ \vol{S_{i}} \geq (2/17) \vol{V}$, and
\item [(c)] $\vol{S_{i} \intersect V_{i}} \leq (1/2) \vol{V_{i}}$, 
\end{itemize}
 then
\[
\conduc{G \{V_{i}  \} }{S_{i} \intersect V_{i}}
\leq 2 \epsilon f_{1} (\phi /104).
\]
Moreover, if $\vol{S_{i} \intersect V_{i} \intersect D_{i+1}} 
  \geq (1- 2 \epsilon ) \vol{S_{i} \intersect V_{i}}$
then
\[
  \vol{D_{i+1}} \geq (1/29) \vol{V}.
\]
\end{lemma}
\begin{proof}
We first lower-bound the volume of the 
  intersection of $S_{i}$ with $V_{i}$ by
\[
  \vol{S_{i} \intersect V_{i}} \geq \vol{S_{i}} - (\vol{V} - \vol{V_{i}})
  \geq \vol{S_{i}} - (1/17) \vol{V} \geq (1/2) \vol{S_{i}}.
\]
We then apply Claim~\ref{clm:ac0} to show
\[
\conduc{G \{V_{i}  \}}{S_{i} \intersect V_{i}}
=
\frac{
  \sizeof{\bdry{V_{i}}{S_{i} \intersect V_{i}}}
}{
  \vol{S_{i} \intersect V_{i}}
}
\leq 
\frac{
  \sizeof{\bdry{W_{i}}{S_{i}}}
}{
  (1/2) \vol{S_{i}}
}
\leq 2 \sigma_{i}
\leq 2 \epsilon f_{1} (\phi /104).
\]
The last part of the lemma follows from 
  $\vol{S_{i} \intersect V_{i}} \geq (1/17) \vol{V}$ 
  and $\epsilon \leq 1/5$.  
\end{proof}

\begin{lemma}\label{lem:ac6}
If
\begin{itemize}
\item [(a)] $\vol{V_{i}} \geq (16/17) \vol{V}$  and
\item [(b)] $\vol{S_{i} \intersect V_{i}} \geq (1/2) \vol{V_{i}}$,
\end{itemize}
 then
\[
\conduc{G \{V_{i}  \}}{S_{i} \intersect V_{i}}
\leq 2 \epsilon f_{1} (\phi /104).
\]
Moreover, if $\vol{(V_{i} - (S_{i} \intersect V_{i})) \intersect D_{i+1}} 
  \geq (1-\epsilon ) \vol{(V_{i} - (S_{i} \intersect V_{i}))}$
then
\[
  \vol{D_{i+1}} \geq (3/16) \vol{V}.
\]
\end{lemma}
\begin{proof}
As 
  $\vol{S_{i}} \leq (1/2) \vol{W_{i}} \leq (1/2) \vol{V}$ and
 $\vol{V_{i} - S_{i} \intersect V_{i}}
  \geq \vol{V_{i}} - \vol{S_{i}} 
  \geq (15/34) \vol{V}$,
\[
  \vol{V_{i} - S_{i} \intersect V_{i}} \geq (15/17) \vol{S_{i}}.
\]
So, by Claim~\ref{clm:ac0},
\[
 \conduc{G \{V_{i} \}}{V_{i} - (V_{i} \intersect S_{i})} 
=
\frac{
 \sizeof{\bdry{V_{i}}{S_{i} \intersect V_{i}}}
}{
  \vol{V_{i} - (V_{i} \intersect S_{i})}
}
\leq 
(17/15)
\frac{
  \sizeof{\bdry{W_{i}}{S_{i}}}
}{
  \vol{S_{i}}
}
\leq
(17/15) \sigma_{0}
\leq 
2 \epsilon f_{1} (\phi /104).
\]
The last part now follows from
\[
  \vol{V_{i} - S_{i} \intersect V_{i}} \geq (15/17) \vol{S_{i}}
 \geq \frac{15}{17} \frac{1}{2} \vol{V_{i}} \geq (5/16) \vol{V}
\]
and $\epsilon \leq 1/5$.
\end{proof}

\begin{proof}[Proof of Theorem~\ref{thm:ApproxCut}]
The proofs of (A.1) and (A.2) are similar to the proofs of (Q.1) and (Q.2).

To prove (A.3), 
  we will assume that for each set $S_{i}$ that satisfies conditions
  \eqref{eqn:Q3} in $G \{V_{i} \}$ the call to \partition2 \ succeeds
 and that the same holds for all sets $V_{i} - S_{i}$ that satisfy conditions
  \eqref{eqn:Q3} in $G \{V_{i} \}$.
As this assumption involves at most $2r$ sets, by Lemma~\ref{lem:partition2} 
  it holds with probability at least $1 - 2r (p/2r) = 1-p$.

If there is an $i$ for which $\vol{V_{i}} < (16/17) \vol{V}$,
  then $\vol{D} \geq (1/17) V$
  and condition $(A.3.a)$ is satisfied.
So, we assume that $\vol{V_{i}} \geq (16/17) \vol{V}$ for the rest of the proof.

Observe that the algorithm \approxcut \ calls \partition2 with 
\[
\theta = (2/23) \phi ,
\]
and  that 
\[
  \phi /104 < \theta /9.
\]
So, if $\vol{S_{i} \intersect V_{i}} \leq \vol{V_{i}}/2$ and
\[
  \conduc{G \{ V_{i} \}}{S_{i}} \leq f_{1} (\phi /104),
\]
then $S_{i}$ satisfies the conditions \eqref{eqn:Q3} in $G \{V_{i} \}$.

If there is an $i$ for which $\vol{S_{i}} \geq (2/17) \vol{V}$,
  then by Lemmas~\ref{lem:ac5} and~\ref{lem:ac6} either $S_{i} \intersect V_{i}$
  or $V_{i} - (S_{i} \intersect V_{i})$ satisfies conditions
  \eqref{eqn:Q3} in $G \{V_{i} \}$
 and the success of the call to \partition2 implies
\[
\vol{D} \geq (1/29) \vol{V}.
\]

So, for the rest of the proof we may assume
  $\vol{S_{i}} \leq (2/17) \vol{V}$.
In this case we may show that
\begin{equation}\label{eqn:ac}
  \vol{S_{i} \intersect V_{i+1}} \leq \epsilon \vol{S_{i}}
\end{equation}
as follows.
If $\vol{S_{i} \intersect V_{i}} \leq \epsilon \vol{S_{i}}$ then
  \eqref{eqn:ac} trivially holds.
Otherwise, Lemma~\ref{lem:ac4} tells us that $S_{i}$
  satisfies conditions \eqref{eqn:Q3} in $G \{V_{i} \}$ and
  that the success of the call to \partition2 guarantees \eqref{eqn:ac}.

We may now apply Lemma~\ref{lem:ac2} to show that $S_{r}$ is empty if it
  is defined. 
So, there is an $i$ for which $W_{i} = U_{i}$ and by Claim~\ref{clm:uiExpander}  and
  Lemma~\ref{lem:ac3} 
\[
  \Conducin{W_{i}} \geq \frac{c_{2} \phi^{2}}{\log^{4} m}.
\]
as $V-D = V_{r} \subseteq V_{i} \subseteq W_{i}$, the set $W = W_{i}$
  satisfies (A.3.b).
\end{proof}

\section{Sparsifying Unweighted Graphs}\label{sec:unweighted}
We now show how to use the algorithms \approxcut\ and \sample\
  to sparsify unweighted graphs.
More precisely, we treat every edge in an unweighted graph as an
  edge of weight $1$.
The algorithm \unwtedsparsify \ 
  follows the outline described in Section~\ref{sec:existence}.
Its main subroutine \partsample \ calls \approxcut \ to partition the graph.
Whenever \approxcut \ returns a small cut, we know that the complement
  is contained in a subgraph of large conductance.
In this case, \partsample \ calls \sample \ to sparsify the large part.
Whenever the cut returned by \approxcut \ is large, \partsample \
  recursively acts on the cut and its complement so that it eventually
  partitions and samples both.
The output of \partsample \ is 
  the result of running \sample \ on the graphs
  induced on the 
  vertex sets of a decomposition of the original graph.
The main routine \unwtedsparsify \ calls \partsample \ and then
  acts recursively to sparsify the edges that go between the parts
  of the decomposition produced by \partsample .

\vskip 0.2in
\noindent
\fbox{
\begin{minipage}{6in}
\noindent $\Gtil  = \unwtedsparsify (G, \epsilon , p)$
\begin{enumerate}
\item [1.] If $\vol{V} \leq c_{3} \epsilon^{-2}n \log^{30} (n/p)$, return $G$
 (where $c_{3}$ is set in the proof of Lemma~\ref{lem:partsample}).

\item [2.] Set $\phi = \left(2 \log_{29/28} \vol{V} \right)^{-1}$,
$\phat = p/6n\log_{2}n$, and
 $\epshat = \frac{\epsilon (\ln 2)^{2}}{(1 + 2 \log_{29/28} n) (2   \log  n )}$.

\item [3.] Set $(\Gtil_{1}, \dotsc , \Gtil_{k}) =
   \partsample (G, \phi ,  \epshat, \phat  )$.

\item [4.] Let $V_{1}, \dotsc , V_{k}$ be the vertex sets of
  $\Gtil_{1}, \dotsc , \Gtil_{k}$, respectively, and let
  $G_{0}$ be the graph with vertex set $V$ and edge set
  $\bdry{}{V_{1}, \dotsc , V_{k}}$.

\item [5.] Set $\Gtil_{0} =
  \unwtedsparsify (G_{0}, \epsilon , p)$.

\item [6.] Set $\Gtil = \sum_{i=0}^{k} \Gtil_{i}$.
\end{enumerate}

\noindent $(\Gtil_{1}, \dotsc , \Gtil_{k}) = \partsample  (G = (V,E), \phi , \epshat , \phat  )$
\begin{enumerate}
\item [0.] Set $\lambda = f_{2} (\phi)^{2}/2$, where $f_{2}$ is defined in \eqref{eqn:f4}.
\item [1.] Set $D = \approxcut (G, \phi , \phat )$.
\item [2.] If $D = \emptyset$, return
  $\Gtil_{1} = \sample (G, \epshat , \phat , \lambda )$.

\item [3.] Else, if $\vol{D} \leq (1/29) \vol{V}$
\begin{itemize}
\item [a.] Set $\Gtil_{1}  = \sample (G (V-D), \epshat , \phat , \lambda )$
\item [b.] Return $(\Gtil_{1}, \partsample  (G (D), \phi , \epshat , \phat   ))$.
\end{itemize}

\item [4.] Else,
\begin{itemize}
\item [a.] Set $\Htil_{1}, \dotsc , \Htil_{k} = \partsample  (G (V-D), \phi , \epshat , \phat   )$.
\item [b.] Set $\Itil_{1}, \dotsc , \Itil_{j} = \partsample  (G (D), \phi , \epshat , \phat  )$.
\item [c.] Return $(\Htil_{1}, \dotsc , \Htil_{k}, \Itil_{1}, \dotsc , \Itil_{j})$.
\end{itemize}
\end{enumerate}
\end{minipage}
}
\vskip 0.2in

\begin{lemma}[\partsample]\label{lem:partsample}
Let $G = (V,E)$ be a graph.
Let $\Gtil_{1}, \dotsc , \Gtil_{k}$ be the output
  of $\partsample (G, \phi , \epshat , \phat  )$.
Let $V_{1}, \dotsc , V_{k}$ be the vertex sets of
  $\Gtil_{1}, \dotsc , \Gtil_{k}$, respectively, and let
  $G_{0}$ be the graph with vertex set $V$ and edge set
  $\bdry{}{V_{1}, \dotsc , V_{k}}$.

Then,
\begin{itemize}
\item [(PS.1)]
$\sizeof{\bdry{}{V_{1}, \dotsc , V_{k}}} \leq \sizeof{E}/2$.
\end{itemize}
With probability at least $1-3 n \phat $,
\begin{itemize}
\item [(PS.2)]
  the graph
\[
G_{0} + \sum_{i=1}^{k} \Gtil_{i}
\]
is a $(1+\epshat)^{1+\log_{29/28} \vol{V}}$ approximation of $G$, and
\item [(PS.3)]
the total number of edges in $\Gtil_{1}, \dotsc ,\Gtil_{k}$
  is at most $c_{3} \epsilon^{{-2}}\sizeof{V}\log^{30} (n/p)$, for some 
  absolute constant $c_{3}$.
\end{itemize}

\end{lemma}
\begin{proof}
We first observe that whenever the algorithm calls itself recursively, the volume
  of the graph in the recursive call is at most $28/29$ of the volume of the
  input graph.
So, the recursion depth of the algorithm is at most $\log_{29/28} \vol{V}$.
Property $(PS.1)$ is a consequence of part $(A.2)$ of Theorem~\ref{thm:ApproxCut}
  and this bound on the recursion depth.

We will assume for the rest of the analysis that
\begin{enumerate}
\item [1.] for every call to \sample \ in line 2,
  $\Gtil_{1}$ is a $(1+\epshat )$ approximation of $G$ and the number
  of edges in $\Gtil_{1}$ satisfies (S.2),

\item [2.] for every call to \sample \ in line 3a,
  $\Gtil_{1} + G (D) + \bdry{}{D,V-D}$ is a
  $(1+\epshat )$ approximation of $G$  and the number
  of edges in $\Gtil_{1}$ satisfies (S.2), and
\item [3.] For every call to \approxcut \ in line 1
  for which the set $D$ returned satisfies $\vol{D} \leq (1/29) \vol{V}$,
  there exists a set $W$ containing $V-D$ for which $\Conducin{W} \geq f_{2} (\phi)$,
  where $f_{2}$ was defined in \eqref{eqn:f4}.
\end{enumerate}
First observe that at most $n$ calls are made to \sample \ and \approxcut \
  during the course of the algorithm.
By Theorem~\ref{thm:ApproxCut}, the probability that assumption $3$ fails is at
  most $n \phat $.
If assumption $3$ never fails, 
  we may apply Theorem~\ref{thm:sampling} to prove that assumptions 1 and 2
  probably hold, as follows.
Consider a subgraph $G (V-D)$ on which \sample \ is called, using $D = \emptyset$
  if \sample \ is called on line 2.
Assumption 3 tells us that there is a set $W \supseteq V - D$
  for which $\Conducin{W} \geq f_{2} (\phi)$.
Theorem~\ref{thm:cheeger} tells us that
  the smallest non-zero normalized Laplacian eigenvalue of $G (W)$ is at least
  $\lambda$, where $\lambda$ is set in line $0$.
Treating $G (W)$ as the input graph, and $S = V - D$, 
  we may apply Theorem~\ref{thm:sampling} to show that
  assumptions $1$ and $2$ fail with probability at most $\phat $ each.
Thus, all three assumptions hold with probability at least $1 - 3 n \phat $.

Property $(PS.3)$, and the existence of the constant $c_{3}$,
   is a consequence of assumptions $1$ and $2$.
Using these assumptions, we will now establish $(PS.2)$ by induction on
  the depth of the recursion.
For a graph $G$ on which \partsample \ is called, let $d$ be the maximum
  depth of recursive calls of the algorithm on $G$, let
 $\Gtil_{1}, \dotsc , \Gtil_{k}$ be output of \partsample \ on $G$,
  and let $V_{1}, \dotsc , V_{k}$ be the vertex sets of
  $\Gtil_{1}, \dotsc , \Gtil_{k}$, respectively.
We will prove by induction on $d$ that
\begin{equation}\label{eqn:partsparse}
\text{
 $\sum_{i=1}^{k} \Gtil_{i} + \bdry{}{V_{1}, \dotsc , V_{k}}$
is a $(1+\epshat)^{d+1}$-approximation of $G$.}
\end{equation}

We base our induction on the case in which the algorithm does not call itself,
  in which case it returns
  the output of \sample \ in line $2$, and
  the assertion follows from
  assumption 1.

Let $D$ be the set of vertices returned by \approxcut.
If $D \not = \emptyset$, then $d \geq 1$.
We first consider the case in which
  $\vol{D} \leq  (1/29) \vol{V}$.
In this case, let
  $H = G (D)$,
 let  $\Htil_{1}, \dotsc , \Htil_{k}$
  be the graphs returned by
  the recursive call to $\partsample$ on $H$,
  and let $W_{1}, \dotsc , W_{k}$ be the vertex sets of $\Htil_{1}, \dotsc , \Htil_{k}$.
Let $H_{0}$ be the graph on vertex set $D$ with edges $\bdry{}{W_{1}, \dotsc , W_{k}}$.
We may assume by way of induction that
\[
  H_{0} + \sum_{i=1}^{k} \Htil_{i}
\]
is a $(1+\epshat)^{d}$-approximation of $H$.
We then have
\begin{align*}
  G
& = G (V-D) + H + \bdry{}{V-D,D}\\
& \pleq (1+\epshat) \left(\Gtil_{1} + H +  \bdry{}{V-D,D} \right),
    & \text{by assumption 2,}\\
& \pleq (1+\epshat) \left(\Gtil_{1} +
  (1+\epshat)^{d}\left(\sum_{i=1}^{k}\Htil_{i} + H_{0} \right) + \bdry{}{V-D,D}   \right),
& \text{by induction,}
\\
& \pleq
  (1+\epshat)^{d+1} \left(\Gtil_{1} + \sum_{i=1}^{k}\Htil_{i} + H_{0} + \bdry{}{V-D,D} \right)\\
& =
  (1+\epshat)^{d+1} \left(\Gtil_{1} + \sum_{i=1}^{k}\Htil_{i} +
  \bdry{}{V-D, W_{1}, \dotsc , W_{k}}   \right).
\end{align*}
One may similarly prove
\[
(1+\epshat)^{d+1} G \pgeq \left(\Gtil_{1} + \sum_{i=1}^{k}\Htil_{i} +
  \bdry{}{V-D, W_{1}, \dotsc , W_{k}}   \right),
\]
establishing \eqref{eqn:partsparse} for $G$.

We now consider the case in which
  $\vol{D} > (1/29) \vol{V}$.
In this case, let
  $H = G (D)$ and $I = G (V-D)$.
Let $W_{1}, \dotsc , W_{k}$ be the vertex sets of $\Htil_{1}, \dotsc , \Htil_{k}$
  and let $U_{1}, \dotsc , U_{j}$
  be the vertex sets of $\Itil_{1}, \dotsc \Itil_{j}$.
By our inductive hypothesis, we may assume that
$\bdry{}{W_{1}, \dotsc , W_{j}} + \sum_{i=1}^{k} \Htil_{i}$
  is a $(1+\epshat)^{d}$-approximation of $H$ and that
$\bdry{}{U_{1}, \dotsc , U_{j}} + \sum_{i=1}^{j} \Itil_{i}$
  is a $(1+\epshat)^{d}$-approximation of $I$.
These two assumptions immediately imply that
\[
\bdry{}{W_{1}, \dotsc , W_{j}, U_{1}, \dotsc , U_{j}} + \sum_{i=1}^{k} \Htil_{i}
 + \sum_{i=1}^{j} \Itil_{i}
\]
is a $(1+\epshat)^{d}$-approximation of $G$,
  establishing \eqref{eqn:partsparse} in the second case.

As the recursion depth of this algorithm is bounded by
  $\log_{29/28} \vol{V}$, we have established property $(PS.2)$.
\end{proof}

\begin{lemma}[\unwtedsparsify]\label{lem:unwtsparsify}
For $\epsilon , p \in (0,1/2)$ and an unweighted graph $G$ with $n$ vertices, let
$\Gtil$ be the output of $\unwtedsparsify (G, \epsilon , p)$.
Then,
\begin{itemize}
\item [(U.1)] The edges of $\Gtil$ are a subset of the edges of $G$; and
\end{itemize}
 with probability at least $1-p$,
\begin{itemize}
\item [(U.2)] $\Gtil$ is a $(1+\epsilon)$-approximation of $G$, and
\item [(U.3)] $\Gtil$ has at most $c_{4} \epsilon^{-2}n \log^{31} (n/p)$
  edges, for some constant $c_{4}$.
\end{itemize}
\end{lemma}
Moreover, the expected running time of \unwtedsparsify \ is 
  $O \left(m \log (1/p) \log^{15} n \right) $.

\begin{proof}
From $(PS.1)$, we know that the depth of the recursion of
  \unwtedsparsify \ on $G$ is at most $\log_{2} \vol{V} \leq 2 \log n$.
So, with probability at least
\[
  1 - (2 \log n) \cdot 3 n \phat = 1- p,
\]
properties $(PS.2)$ and $(PS.3)$ hold for the output of
  \partsample \ every time it is called by \unwtedsparsify.
For the rest of the proof, we assume that this is the case.

Claim $(U.3)$ follows immediately from $(PS.3)$ and the bound on the
  recursion depth of \unwtedsparsify .
We prove claim $(U.2)$ by induction on the recursion depth.
In particular, we prove that if \unwtedsparsify \ makes $d$ recursive
  calls to itself on graph $G$, then the graph $\Gtil$ returned is
  a $(1+\epsilon \ln 2 / (2 \log n + 1))^{d}$ approximation of $G$.
We base the induction in the case where \unwtedsparsify \ makes no
  recursive calls to itself, in which case it returns at line 1
  with a $1$-approximation.

For $d > 0$, we assume for induction that $\Gtil_{0}$ is a
  $(1+\epsilon \ln 2 / 2 \log n)^{d-1}$-approximation of $G_{0}$.
By the assumption that $(PS.2)$ holds, we know that
  $G_{0} + \sum_{i=1}^{k} \Gtil_{i}$
  is a
\[
  (1 + \epshat)^{(1 + \log_{29/28}  n^{2})}
 \leq
  (1 + \epsilon \ln 2/ (2 \log n ))
\]
approximation of $G$, as $\epsilon \ln 2 / (2 \log n) \leq 1$
  (here, we apply the inequality $(1 + x \ln 2 / k)^{k} \leq 1+x$).
By following the arithmetic in the proof of Lemma~\ref{lem:partsample},
  we may prove that
$\Gtil _{0} + \sum_{i=1}^{k} \Gtil_{i}$
  is a $(1+\epsilon \ln 2/ (2 \log n))^{d}$ approximation of $G$.

To finish, we observe that
\[
(1+\epsilon \ln 2 / (2 \log n ))^{2 \log n} \leq 1 + \epsilon ,
\]
for $\epsilon < 1$.

Claim $(U.1)$ follows from the observation that the set of edges of
  the graph output by \sample \ is a subset of the set of edges of its input.

To bound the expected running time of \unwtedsparsify,
  observe that the bound on the recursion depth of
  \partsample \ implies that its expected running time is at most
  $O (\log n)$ times the expected running time
  of \approxcut \ with $\phi = \Omega(1/\log n)$,
  plus the time required to make the calls
  to sample, which is at most $O (m)$.

Another multiplicative factor of $O (\log n)$ comes from the
  logarithmic number of times that \unwtedsparsify \ can call
  itself during the recursion.
\end{proof}

\section{Sparsifying Weighted Graphs}\label{sec:weighted}
In this section, we show how to sparsify graphs whose edges have
  arbitrary weights.
We begin by showing how to sparsify weighted graphs
  whose edge weights are integers in the range
  $\setof{1, \dotsc , U}$.
One may also think of this as sparsifying a multigraph.
This first result will follow simply from the algorithm for
  sparsifying unweighted graphs, at a cost of a $O (\log U)$ factor
   in the number of edges in the sparsifier.

We then explain the obstacle to sparsifying arbitrarily weighted graphs
  and how we overcome it.
We end the section by proving that it is possible to modify our construction
  of sparsifiers
  so that for every node the total blow-up in weight of the edges attached to
  it is bounded.

\subsection{Bounded Weights}

We recall that we treat an unweighted graph as a graph in which every edge has
  weight 1, and for clarity we often refer to such a graph as a
  \textit{weight-1 graph}.
Our algorithm for sparsifying graphs with weights in
  $\setof{1,\dotsc ,U-1}$ works by constructing
  $\log_{2} U$ weight-1 graphs $G_{i}$ and then expressing
  $G$ as a sum of $2^{i} G_{i}$.
Each edge of $G$ appears in the graphs $G_{i}$ for which the $i$th   
  bit of the binary expansion of the weight of the edge is $1$.
We sparsify the graphs $G_{i}$ independently, and then sum the results.

\vskip 0.2in
\noindent
\fbox{
\begin{minipage}{6in}
\noindent $\Gtil  = \boundedsparsify (G, \epsilon , p)$, $G = (V,E,w)$
  has integral weights in $[1, 2^{u})$.
\begin{enumerate}
\item [1.] Decompose $G$ as
\[
  G = \sum_{i = 0}^{u-1} 2^{i} G_{i},
\]
where each $G_{i}$ is a weight-1 graph.

\item [2.] For each $i$, set $\Gtil_{i} = \unwtedsparsify (G_{i}, \epsilon , p/ u)$.
\item [3.] Return $\Gtil = \sum_{i} 2^{i}\Gtil_{i}$.
\end{enumerate}
\end{minipage}
}
\vskip 0.2in

\begin{lemma}[\boundedsparsify]\label{lem:boundedsparsify}
For $\epsilon , p \in (0,1/2)$ and a graph $G$ with integral weights
  and with $n$ vertices, let
$\Gtil$ be the output of $\boundedsparsify (G, \epsilon , p)$.
Let $U-1$ be the maximum weight of an edge in $G$.
Then,
\begin{itemize}
\item [(B.1)] The edges of $\Gtil$ are a subset of the edges of $G$; and,
\end{itemize}
with probability at least $1-p$,
\begin{itemize}
\item [(B.2)] $\Gtil$ is a $(1+\epsilon)$-approximation of $G$, and
\item [(B.3)] $\Gtil$ has at most $c_{4}\epsilon^{{-2}} n \log U \log^{31} (n/p)$
  edges.
\end{itemize}
\end{lemma}
Moreover, the expected running time of \boundedsparsify  \ is 
  $O \left(m \log U \log (1/p) \log^{15} n \right) $.
\begin{proof}
Immediate from Lemma~\ref{lem:unwtsparsify}.
\end{proof}

\subsection{Coping with Arbitrary Weights: Graph Contraction}\label{}

When faced with an arbitrary weighted graph, we will
  first approximate the weight of every edge by the sum of a
  few powers of two.
However, if the weights are arbitrary many different powers
  of two could be required, and we could not construct a sparsifier
  by treating each power of two separately as we did in
  \boundedsparsify .
To get around this problem, we observe that when we are considering
  edges of a given weight, we can assume that all edges of much greater
  weight have been contracted.
We formalize this idea in Lemma~\ref{lem:pullback}.

By exploiting this idea, we are able to
  sparsify arbitrary weighted graphs with at most a $O (\log (1/\epsilon))$-factor
  more edges than employed in \boundedsparsify \ when $U = n$.
Our technique is inspired by how Benczur and Karger~\cite{BenczurKarger}
  built cut sparsifiers for weighted graphs out of
  cut sparsifiers for unweighted graphs.

Given a weighted graph $G = (V, E, w)$ and a partition $V_{1}, \dotsc , V_{k}$ of V,
  we define the \textit{map} of the partition to be the function
\[
 \pi : V \rightarrow \setof{1,\dotsc ,k}
\]
for which $\pi (u) = i$ if $u \in V_{i}$.
We define the \textit{contraction} of $G$ under $\pi$ to be the weighted graph
  $H = (\setof{1,\dotsc ,k}, F, z)$, where $F$ consists of edges of the form
  $(\pi (u), \pi (v))$ for $(u,v) \in E$, and where the weight of edge
  $(i,j) \in F$ is
\[
z (i,j) =   \sum_{(u,v) : \pi (u) = i, \pi (v) = j} w (u,v).
\]
We do not include self-loops in the contraction, so
  edges $(u,v) \in E$ for which $\pi (u) = \pi (v)$ do not appear in the contraction.

Given a weighted graph $\Htil = (\setof{1,\dotsc ,k}, \Ftil , \ztil)$,
  we say that $\Gtil = (V, \Etil , \wtil)$
  is a \textit{pullback} of $\Htil$ under $\pi$ if
\begin{itemize}
\item [1.]  $\Htil$ is the contraction of $\Gtil$ under $\pi$, and
\item [2.]  for every edge
  $(i,j) \in \Ftil$, $\Etil$ contains exactly one edge $(u,v)$
  for which $\pi (u) = i$ and $\pi (v) = j$.
\end{itemize}

In the following lemma, we consider a graph in which each of the vertex
  sets $V_{1},\dotsc , V_{k}$ are connected by edges of high weight while
  all the edges that go between these sets have low weight.
We show that one can sparsify the low-weight edges by taking a pullback
  of an approximation of the contraction of the graph.

\begin{lemma}[Pullback]\label{lem:pullback}
Let $G = (V, E, w)$ be a weighted graph, let $V_{1}, \dotsc , V_{k}$
  be a partition of $V$, and let $\pi$ be the map of the partition.
Set $E_{0} = \bdry{}{V_{1}, \dotsc , V_{k}}$,
  $G_{0} = (V, E_{0}, w)$,
  $E_{1} = E - E_{0}$, and
  $G_{1} = (V, E_{1}, w)$.
For some $\epsilon < 1/2$ let
 $\Gtil_{0} $ be a pullback under $\pi$ of a
  $(1+\epsilon)$-approximation of the contraction of $G_{0}$ under $\pi$.
Assuming that $c \geq 3$,
\begin{itemize}
\item [1.] each set of vertices $V_{i}$ is connected by edges in $E_{1}$,
\item [2.] every edge in $E_{1}$ has weight at least $c^{2} n^{3}$, and
\item [3.] every edge in $E_{0}$ has weight 1.
\end{itemize}
Then, $\Gtil_{0} + G_{1}$ is an $\alpha$-approximation of $G$,
  for
\[
 \alpha  = (1+\epsilon) (1+1/c)^{2}.
\]
\end{lemma}

Our proof of Lemma~\ref{lem:pullback} uses the following lemma bounding
  how well a path preconditions an edge.
It is an example of a Poincar{\'e} inequality~\cite{DiaconisStrook},
  and it may be derived from  the
  Rank-One Support Lemma of~\cite{SupportTheory},
  the Congestion-Dilation Lemma of~\cite{SupportGraph},
  or the Path Lemma of~\cite{SpielmanTengLinsolve}.
We include a proof for convenience.
\begin{lemma}\label{lem:Path}
Let $(u,v)$ be an edge of weight $1$,
  and let $F$ consist of a path from $u$ to $v$ in which
  the edges on the path have weights
  $w_{1}, \dotsc , w_{k}$.
Then,
\[
(u,v) \pleq \left(1/w_{1} + \dotsb + 1/w_{k} \right) F.
\]
\end{lemma}
\begin{proof}
Name the vertices on the path $0$ through $k$ with vertex $0$
  replacing $u$ and vertex $k$ replacing $v$.
Let $w_{i}$ denote the weight of edge $(i,i-1)$.
We need to prove that for every vector $x$,
\[
  \left(x (k) - x (0) \right)^{2} \leq 
  \left(\sum_{i=1}^{k} \frac{1}{w_{i}} \right) 
  \sum_{i=1}^{k} w_{i} (x (i) - x (i-1))^{2}.
\]
For $1 \leq i \leq k$ set $y (i) = \sqrt{w_{i}} (x_{i}-x_{i-1}) $.
The Cauchy-Schwarz inequality now tells us that
\[
    \left(x (k) - x (0) \right)^{2}
= 
\left(\sum_{i=1}^{k} \sqrt{w_{i}} (x_{i}-x_{i-1}) / \sqrt{w_{i}} \right)^{2}
\leq 
\left(\sum_{i=1}^{k} \left(1 / \sqrt{w_{i}} \right)^{2} \right)
\left(\sum_{i=1}^{k} \left(\sqrt{w_{i}} (x_{i}-x_{i-1}) \right)^{2}
\right),
\]
as required.
\end{proof}

\begin{proof}[Proof of Lemma~\ref{lem:pullback}]
Let $H$ be the contraction of $G_{0}$ under $\pi$,
  and let $\Htil$ be the $(1+\epsilon)$-approximation of $H$
  for which $\Gtil_{0}$ is a pullback.

We begin the proof by choosing an arbitrary vertex $v_{i}$ in each set $V_{i}$.
Now, let $F$ be the weighted
  graph on vertex set $\setof{v_{1}, \dotsc , v_{k}}$
  isomorphic to $H$ under the map $i \mapsto v_{i}$,
 and let $\Ftil$ be the analogous graph for $\Htil$.
Our analysis will go through an examination of the graphs
\[
  I \defeq F + G_{1} \quad \text{and} \quad
  \Itil \defeq \Ftil + G_{1}.
\]
The lemma is a consequence of the following three statements,
  which we will prove momentarily:
\begin{itemize}
\item [(a)] $I$ is a $(1+1/c)$-approximation of $G$.
\item [(b)] $\Itil$ is a $(1+\epsilon)$-approximation of $I$.
\item [(c)] $\Itil$ is a $(1+1/c)$-approximation of $\Gtil_{0} + G_{1}$.
\end{itemize}
To prove claim (a), consider any edge $(a,b) \in E_{0}$.
As $\pi (a) \not = \pi (b)$, the graph
  $\frac{1}{c n^{2}}G_{1}$ contains a path from
  $a$ to $v_{\pi (a)}$ and a path from $b$ to $v_{\pi (b)}$.
The sum of the lengths of these paths is at most $n$, and each
  edge on each path has weight at least $c n$.
So, if we let $f$ denote an edge of weight $1$ from
  $\pi (a)$ to $\pi (b)$,
  then Lemma~\ref{lem:Path} tells us that
\begin{equation}\label{eqn:pullback1n}
(a,b) \pleq (1/1 + n / c n) \left(f + \frac{1}{c n^{2}} G_{1} \right)
=
(1+1/c) \left(f + \frac{1}{c n^{2}} G_{1} \right),
\end{equation}
and
\begin{equation}\label{eqn:pullback2}
f \pleq (1+1/c) \left((a,b) + \frac{1}{c n^{2}} G_{1} \right).
\end{equation}
As there are fewer than $n^{2}/2$ edges in $E_{0}$, we may
 sum \eqref{eqn:pullback1n} over all of them 
  to establish
\[
  G_{0}  \pleq (1+1/c) \left[F + \frac{1}{2c} G_{1} \right].
\]
So,
\begin{align*}
  G_{0} + G_{1}
&  \pleq (1+1/c) \left[F + \frac{1}{2c} G_{1} \right] + G_{1}\\
&  \pleq (1+1/c) \left[F + G_{1} \right],
\end{align*}
as $c \geq 1$.
The inequality
\[
F + G_{1} \pleq (1+1/c)\left[ G_{0} + G_{1} \right],
\]
and thus part $(a)$, may be established by similarly summing
  over inequality \eqref{eqn:pullback2}.

Part $(b)$ is immediate from the facts that
   $\Ftil$ is a $(1+\epsilon)$-approximation
  of $F$, that $I = F + G_{1}$ and $\Itil = \Ftil + G_{1}$.

Part $(c)$ is very similar to part $(a)$.
We first note that the sum of the weights of edges in $\Ftil$
  is at most $(1+\epsilon)$ times the sum of the weights of edges
  in $F$, and so is at most $(1+\epsilon) n^{2}/2$.
Now, for each edge $(a,b)$ in $\Gtil_{0}$ of weight $w$, there is
  a corresponding edge $(v_{\pi (a)}, v_{\pi (b)})$ of weight $w$
  in $\Ftil$.
Let $e$ denote the edge $(a,b)$ of weight $w$ and let $f$ denote
  the edge $(v_{\pi (a)}, v_{\pi (b)})$ of weight $w$.
As in the proof of part $(a)$, we have
\[
  e \pleq (1+1/c) \left(f + \frac{w}{c n^{2}} G_{1}  \right),
\]
and
\[
  f \pleq (1+1/c) \left(e + \frac{w}{c n^{2}} G_{1}  \right).
\]
Summing these inequalities over all edges in $\Etil_{0}$,
  adding $G_{1}$ to each side, and recalling
  $\epsilon \leq 1/2$ and $c \geq 3$,
  we establish part $(c)$.
\end{proof}

We now state the algorithm \sparsify .
For simplicity of exposition, 
  we assume that the weights of edges in its input are all
  at most $1$.
However, this is not a restriction as one can scale down
   the weights of any graph to satisfy this requirement, apply \sparsify ,
  and then scale back up.

The algorithm \sparsify \ first replaces each weight $w_{e}$ with
  its truncation to its few most significant bits, $z_{e}$.
The resulting modified graph is called $\Ghat$.
As $z_{e}$ is very close to $w_{e}$, little is lost by this substitution.
As in \boundedsparsify , $\Ghat$ is represented as a sum of graphs 
  $2^{-i} G^{i}$ where each $G^{i}$ is a weight-1 graph.
Because the weight of every edge in $\Ghat$ only has a few bits,
  each edge only appears in a few of the graphs $G^{i}$.

Our first instinct would be to sparsify each of the graphs $G^{i}$ individually.
However, this could result in too many edges as sparsifying produces
  a graph whose number of edges is proportional to its number of vertices,
  and the sum over $i$ of the number of vertices in each $G^{i}$ could be large.
To get around this problem, we contract all edges of much higher weight
  before sparsifying.
In particular, the algorithm \sparsify \
  partitions the vertices into components that are connected by
  edges of much higher weight.
It then replaces each $G^{i}$ with a pullback of a
  sparsifier of the contraction of $G^{i}$ under this partition.
In Lemma~\ref{lem:sparsifyNumClusters} we prove that the sum over $i$ of the
  number of vertices in the contraction of each $G^{i}$ will only be
  a small multiple of $n$.

\noindent
\fbox{
\begin{minipage}{6in}
\noindent $\Gtil  = \sparsify (G, \epsilon , p)$,
where $G = (V,E,w)$ and $w (e) \leq 1$ for all $e \in E$.
\begin{enumerate}

\item [0.] Set $Q = \ceiling{6/\epsilon }$,
  $b = 6 / \epsilon $,
  $c = 6 /\epsilon $,
  $\epshat = \epsilon / 6$, and
  $l = \ceiling{\log_{2} 2 b c^{2} n^{3}}$.

\item [1.] For each edge $e \in E$,
\begin{enumerate}
\item [a.] choose
  $r_{e}$ so that $Q \leq 2^{r_{e}} w_{e} < 2 Q$,
\item [b.] let $q_{e}$ be the largest integer such that
  $q_{e} 2^{-r_{e}} \leq w_{e}$, (and note $Q \leq q_{e} < 2Q$)
\item [c.] set $z_{e} = q_{e} 2^{-r_{e}}$.
\end{enumerate}

\item [2.] Let $\Ghat = (V,E,z)$, and
  express
\[
  \Ghat = \sum_{i \geq 0} 2^{-i} G^{i},
\]
where in each graph $G^{i}$ all edges have weight $1$,
  and each edge appears in at most $\ceiling{\log_{2} 2 Q}$
  of these graphs.

\item [3.]
  Let $E^{i}$ be the edge set of $G^{i}$.
  Let $E^{\leq i} = \union_{j \leq i} E^{j}$.
  For each $i$, let $D^{\leq i}_{1}, \dotsc , D^{\leq i}_{\eta_{i}}$
  be the connected components of $V$ under $E^{\leq i}$.
For $i = 0$, set $\eta_{i} = 0$.

\item [4.] For each $i$ for which $E^{i}$ is non-empty,
\begin{enumerate}
\item [a.] Let $V^{i}$ be the set of vertices attached to edges
  in $E^{i}$.

\item [b.]
  Let $C^{i}_{1}, \dotsc , C^{i}_{k_{i}}$ be the sets of form
  $D^{\leq i - l}_{j} \intersect V^{i}$ that are non-empty
  and have an edge of $E^{i}$ on their boundary,
  (that is, the interesting components of $V^{i}$ after contracting edges
  in $E^{\leq i-l}$).  Let $W^{i} = \union_{j} C^{i}_{j}$.

\item [c.] Let $\pi$ be the map of partition $C^{i}_{1},\dotsc ,C^{i}_{k_{i}}$,
  and let $H^{i}$ be the contraction of $(W^{i}, E^{i})$ under $\pi$.

\item [d.] $\Htil^{i} = \boundedsparsify (H^{i}, \epshat , p/ (2nl))$.
\item [e.] Let $\Gtil^{i}$ be a pullback of $\Htil^{i}$ under $\pi$ whose
  edges are a subset of $E^{i}$.

\end{enumerate}
\item [5.] Return $\Gtil = \sum_{i}  2^{-i} \Gtil^{i}$.
\end{enumerate}
\end{minipage}
}
\vskip 0.125in

\begin{lemma}\label{lem:sparsifyNumClusters}
Let $k_{i}$ denote the number of clusters described by \sparsify \
  at step 4b.
Then,
\[
  \sum_{i} k_{i} \leq  2 nl.
\]
\end{lemma}
\begin{proof}
Let $\eta_{i}$ denote the number of connected components
  in the graph $(V,E^{\leq i})$.
Each cluster $C^{i}_{j}$ has at least one edge of $E^{i}$
  leaving it.
As each pair of components under $E^{\leq i - l}$ that are joined by
  an edge of $E^{i}$ appear in the same component under $E^{\leq i}$,
\[
  \eta_{i} \leq \eta_{i-l} - k_{i}/2.
\]
As the number of clusters never goes negative and is initially at
  most $n$, we may conclude
\[
  \sum_{i} k_{i} \leq 2 nl.
\]
\end{proof}

\begin{theorem}[\sparsify]\label{thm:sparsify}
For $\epsilon \in (1/n, 1/3)$, $p \in (0,1/2)$ and a weighted graph $G$
  and with $n$ vertices in which every edge has weight at most 1.
Let $\Gtil$ be the output of $\sparsify (G, \epsilon , p)$.
\begin{itemize}
\item [(X.1)] The edges of $\Gtil$ are a subset of the edges of $G$; and
\end{itemize}
with probability at least $1-p$,
\begin{itemize}
\item [(X.2)] $\Gtil$ is a $(1+\epsilon)$-approximation of $G$, and
\item [(X.3)] $\Gtil$ has at most
  $c_{5}  \epsilon^{-2}n \log^{33} (n/p)$
  edges, for some constant $c_{5}$.
\end{itemize}
Moreover, the expected running time of \sparsify  \ is 
  $O \left(m \log (1/p) \log^{17} n \right) $.
\end{theorem}
\begin{proof}
To establish property $(X.1)$, it suffices to show that step 4e can
  actually be implemented.
That is, we need to know that all edges in $\Htil^{i}$ can be pulled
  back to edges of $E^{i}$.
This follows from $(B.1)$ and the fact that $H^{i}$ is a contraction
  of $E^{i}$.

We now establish that the graph $\Ghat$ is a $(1+1/Q)$-approximation
  of $G$.
We will then spend the rest of the proof establishing that $\Gtil$
  approximates $\Ghat$.
As the weight of every edge in $\Ghat$ is less than the corresponding
  weight in $G$, we have $\Ghat \pleq G$.
On the other hand, for every edge $e \in E$, $w_{e} \leq (1+1/Q) z_{e}$,
  so $G \pleq (1+1/Q) \Ghat$, and
  $\Ghat$ is a $(1+1/Q)$-approximation of $G$.

From Lemma~\ref{lem:sparsifyNumClusters}, we know that there are at most
  $nl$ values of $i$ for which $k_{i} \geq 2$,
  and so \boundedsparsify \ is called at most $nl$ times.
Thus, with probability at least $1-p$, the output returned by every call
  to \boundedsparsify \ satisfies properties $(B.2)$ and $(B.3)$, and accordingly
  we will assume that these properties are satisfied for the rest of the proof.

As each edge set $E^{i}$ has at most $n^{2}$ edges, the weight
  of every edge in graph $H^{i}$ is an integer between $1$ and $n^{2}$.
So, by property $(B.3)$, the number of edges in $\Htil_{i}$ , and therefore
  in $\Gtil_{i}$, is at most
\[
c_{4} \epshat^{-2}  k_{i} \log n^{2} \log^{31} (k_{i}/ (p / (2nl))) 
\leq
c_{4} \epshat^{-2}  k_{i} \log^{32} ( n^{2}l/ p).
\]
Applying Lemma~\ref{lem:sparsifyNumClusters}, we may prove that the number of edges
  in $\Gtil$ is at most
\[
\sum_{i}
c_{4} \epshat^{-2} k_{i}  \log^{32} ( n^{2}l/ p) 
\leq
c_{4} \epshat^{-2} (2 nl)  \log^{32} ( n^{2}l/ p)
\leq
c_{5} \epsilon^{-2} n  \log^{33} (n/ p), \quad \text{as $\epsilon > 1/n$,}
\]
for some constant $c_{5}$,
  thereby establishing $(X.3)$.

To establish $(X.2)$,
 define for \textit{every} $i$ the weight-1 graph
  $F^{i} = (V, E^{\leq i})$, and observe that
\[
  \sum_{i \geq 0} 2^{-i} F^{i} =  2 \Ghat.
\]
We may apply $(B.2)$ and Lemma~\ref{lem:pullback} to show that
\[
  \Gtil^{i} + c^{2} n^{3} F^{i - l}
\]
is a $(1+\epshat) (1+1/c)^{2}$-approximation of $G^{i} + c^{2} n^{3} F^{i - l}$.
Summing over $i$ while multiplying the $i$th term by $2^{-i}$, we conclude that
\[
  \sum_{i \geq 0} 2^{-i} \left(\Gtil^{i} + c^{2} n^{3} F^{i - l} \right)
=
  \Gtil + c^{2} n^{3} \sum_{i \geq 0} 2^{-i} F^{i-l}
=
  \Gtil + 2 c^{2} n^{3} 2^{-l} \Ghat
\]
is a $(1+\epshat) (1+1/c)^{2}$-approximation of
\[
 \sum_{i \geq 0} 2^{-i} \left(G^{i} + c^{2} n^{3} F^{i - l} \right)
=
\Ghat  + c^{2} n^{3} \sum_{i} 2^{-i} F^{i-l}
=
\Ghat + 2 c^{2} n^{3} 2^{-l} \Ghat.
\]
Setting
\[
  \beta \defeq 2 c^{2} n^{3} 2^{-l}  \leq  1/b,
\]
we have proved that $\Gtil + \beta  \Ghat $
is a $(1+\epshat) (1+1/c)^{2}$-approximation of
$\left(1 + \beta  \right) \Ghat$,
  and by so
  Proposition~\ref{pro:sparsifyCalc} below,
$\Gtil$
is a
\[
  (1 + \epshat) (1+1/c)^{2} (1+ \beta )
\]
approximation of $\Ghat$.
Property $(X.2)$ now follows from the facts that
   $\Ghat$ is a $(1+1/Q)$-approximation of $G$, and
\[
  (1 + \epshat) (1+1/c)^{2} (1+ \beta ) (1+1/Q)
\leq
  (1 + \epsilon /6)^{5}
\leq
  (1 + \epsilon),
\]
for $\epsilon < 1/2$.

To bound the expected running time of \sparsify,
  we observe that the time of the computation is dominated
  by the calls to \boundedsparsify \ and the time
  required to actually form the graphs $H^{i}$.
The sets $D_{j}^{\leq i}$ may be maintained using
  union-find~\cite{Tarjan}, and so incur a cost of at most $O (n \log n)$
  over the course of the algorithm.
Each graph $H^{i}$ may be formed by determining the component
  of each of its edges, at a cost of $O (\sizeof{E^{i}} \log n)$.
So, the time to form the graphs $H^{i}$ can be bounded by
\[
  O (\sum_{i} \sizeof{E^{i}} \log n) =
  O (m \ceiling{\log 2 Q} \log n)
  = O (m \log (1/\epsilon) \log n).
\]
This is dominated by our upper bound on
  the time required in the calls to \boundedsparsify,
  which is
\[
  O \left(\sum_{i} \sizeof{E^{i}} \log n \lg (1/p) \log^{15} n \right)
=
  O \left(m \log (1/\epsilon ) \log n \lg (1/p) \log^{15} n \right)
=
  O \left(m \log (1/p) \log^{17} n \right).
\]
\end{proof}

\begin{proposition}\label{pro:sparsifyCalc}
If $\beta , \gamma < 1/2$ and
 $\Gtil + \beta \Ghat $ is a $(1+\gamma)$-approximation of $(1+\beta) \Ghat $,
  then
$\Gtil$ is a $(1+\beta) (1+\gamma )$-approximation of $\Ghat $.
\end{proposition}
\begin{proof}
We have
\[
  \Gtil + \beta \Ghat  \pleq (1+\gamma) (1+ \beta) \Ghat ,
\]
which implies
\[
  \Gtil  \pleq (1+\gamma) (1+ \beta) \Ghat .
\]

On the other hand,
\begin{align*}
 (1+ \beta) \Ghat
& \pleq
  (1+\gamma) \left(\Gtil + \beta \Ghat  \right) \quad \text{implies}
\\
(1 - \beta \gamma) \Ghat
 & \pleq
(1+ \gamma) \Gtil , \quad \text{which implies}
\\
\Ghat
 & \pleq
\frac{1+\gamma}{1 - \beta \gamma } \Gtil
\\
&  \pleq
 (1 + \beta) (1 + \gamma ) \Gtil,
\end{align*}
under the conditions $\beta , \gamma < 1/2$.
\end{proof}

\subsection{Bounding Blow-Up}\label{ssec:blowup}
When we approximate a graph $G = (V,E,w)$ by a graph
  $\Gtil = (V, \Etil , \wtil)$ with $\Etil \subseteq E$,
  we define the \textit{blow-up} of an edge $e \in E$
  by
\[
 \blowup{\Gtil }{e} \defeq
\begin{cases}
 \frac{\wtil_{e}}{w_{e}}  & \text{if $e \in \Etil $, and}\\
0 & \text{otherwise}
\end{cases}
\]
Similarly, we define the blow-up of a vertex $v$ to be
\[
  \blowup{\Gtil }{v} \defeq
  \frac{1}{d_{v}} \sum_{(u,v) \in E} \blowup{\Gtil }{(u,v)}.
\]
The algorithm in~\cite{SpielmanTengLinsolve} for solving linear equations
  requires sparsifiers in which every vertex has bounded blow-up.
While the sparsifiers output by \unwtedsparsify \ and \boundedsparsify \
  satisfy this condition with high probability,
  the sparsifiers output by \sparsify  \ do not.
The reason is that nodes of low degree can become part of clusters
  $C^{i}_{j}$ with many edges of $E^{i}$ on their boundary.
These clusters can become vertices of high degree in the contraction by $\pi$,
  and so can become attached to edges of high blow-up when they are sparsified.

This problem may be solved by making two modifications to \sparsify .
First, we sub-divide the clusters $C^{i}_{j}$ so all the vertices in each
  cluster have approximately the same degree, and so that
  the degree of every vertex in $H^{i}$ is at most four times the degree
  of the vertices that map to it.
Then,
  we set $\Gtil_{i}$ to be a \textit{random} pullback of $\Htil_{i}$
  whose edges are a subset of $E$.
That is, for each edge $(c,d) \in \Htil_{i}$ we pull it back to a randomly
  chosen edge $(a,b) \in E$ for which $\pi (a) = c$ and $\pi (b) = d$.
In this way we may guarantee with high probability that no vertex
  has high blow-up.
We now describe the corresponding algorithm \sparsifytwo \ by just listing
  the lines that differ from \sparsify.

\vskip 0.2in
\noindent
\fbox{
\begin{minipage}{6in}
\noindent $\Gtil  = \sparsify2 (G, \epsilon , p)$,
where $G = (V,E,w)$ has all edge-weights at most $1$.
\begin{enumerate}

\item [4a.]
  Let $\preDelta{V}$ be the set of vertices in $V$ with
  degrees in $[2^{\delta}, 2^{\delta +1})$.
  Let $V^{i}$ be the set of vertices attached to edges
  in $E^{i}$.
  Let $\preDelta{V}^{i}$ be the set of vertices in $\preDelta{V} \intersect V^{i}$.

\item [4b.] For each $\delta$,
  let $\preDelta{C}^{i}_{1}, \dotsc , \preDelta{C}^{i}_{k^{\delta}_{i}}$ be the sets of form
  $D^{\leq i - l}_{j} \intersect \preDelta{V}^{i}$ that are non-empty
  and have an edge of $E^{i}$ on their boundary.
  Let $W^{i} = \union_{j, \delta } \preDelta{C}^{i}_{j}$.
  For each set
  $\preDelta{C}^{i}_{j}$ that has
  more than $2^{\delta +2}$
  edges of $E^{i}$ on its boundary,
  sub-divide the set until each part has between $2^{\delta}$ and $2^{\delta +2}$
  edges on its boundary.
  [We will give a procedure to do the subdivision in the paragraph
immediately after this algorithm].
  Let $\preDelta{C}^{i}_{1}, \dotsc , \preDelta{C}^{i}_{t^{\delta}_{i}}$
  be the resulting collection of sets.

\item [4c.] Let $\pi$ be the map of partition
  of $W^{i}$ by the sets $\setof{\preDelta{C}^{i}_{j}}_{j, \delta }$,
  and let $H^{i}$ be the contraction of $(W^{i}, E^{i})$ under $\pi$.

\item [4e.] 
Let $\Htil^{i} = \boundedsparsify (H^{i}, \epshat , p/ (c_{8} n l\log n))$.
Let $\Gtil^{i}$ be a random pullback of $\Htil^{i}$ under $\pi$ whose
  edges are a subset of $E$.

\end{enumerate}
\end{minipage}
}
\vskip 0.2in

We should establish that it is possible to sub-divide the clusters
  as claimed in step 4b.
To see this, recall that each vertex in a set $\preDelta{C}^{i}_{j}$
  has degree at most $2^{\delta +1}$.
So, if we greedily pull off vertices one by one to form a new set, each time we
  move a vertex the boundary of the new set will increase by at most
  $2^{\delta +1}$ and the boundary of the old set will decrease by at most
  $2^{\delta +1}$.
Thus, at the point when the size of the boundary of the new set
  first exceeds $2^{\delta }$, the size of the boundary of the old set must
  be at least $2^{\delta +2} - 2^{\delta} - 2^{\delta +1} \geq 2^{\delta}$.
So, one can perform the subdivision in step 4b by a naive greedy algorithm.

\begin{theorem}[\sparsifytwo]\label{thm:sparsifytwo}
For $\epsilon \in (1/n, 1/3)$, $p \in (0,1/2)$ and a weighted graph $G$
  with $n$ vertices, let
$\Gtil$ be the output of $\sparsifytwo (G, \epsilon , p)$.
Then,
\begin{itemize}
\item [(Y.1)] the edges of $\Gtil$ are a subset of the edges of $G$; and,
\end{itemize}
with probability at least $1-(4/3) p$,
\begin{itemize}
\item [(Y.2)] $\Gtil$ is a $(1+\epsilon)$-approximation of $G$, and
\item [(Y.3)] $\Gtil$ has at most
  $c_{6} \epsilon^{-2} n \log^{34} (n/p)$
  edges, for some constant $c_{6}$,
\item [(Y.4)] every vertex has blow-up at most $2$.
\end{itemize}
Moreover, the expected running time of \sparsifytwo  \ is 
  $O \left(m  \log (1/p) \log^{17} n \right) $.
\end{theorem}

\begin{proof}
To prove $(Y.3)$, we must bound the number of clusters,
  $\sum_{i,\delta }t_{i}^{\delta }$, produced
  in the modified step 4b.
From Lemma~\ref{lem:sparsifyNumClusters}, we know that
\begin{equation}\label{eqn:sumki}
  \sum_{i} k^{\delta}_{i} \leq 2 (l \cdot  n).
\end{equation}
To bound $\sum_{i} t^{\delta}_{i}$, let $\bdry{E_{i}}{W}$ denote the set
  of edges in $E_{i}$ leaving a set of vertices $W$.
Let $S^{\delta}_{i}$ be the set of $j$ for which $\preDelta{C}^{i}_{j}$
  was created by subdivision,
and recall that for all
  $j \in S^{\delta}_{i}$,
\[
  \sizeof{\bdry{E_{i}}{\preDelta{C}^{i}_{j}}} \geq 2^{\delta}.
\]
So,
\[
  \sum_{j \in S^{\delta}_{i}}
    \sizeof{\bdry{E_{i}}{\preDelta{C}^{i}_{j}}}
   \geq 2^{\delta} (t^{\delta}_{i} - k^{\delta}_{i}),
\]
and
\begin{equation}\label{eqn:sparsify2a}
  \sum_{i, j \in S^{\delta}_{i}}
    \sizeof{\bdry{E_{i}}{\preDelta{C}^{i}_{j}}}
   \geq 2^{\delta} \sum_{I} (t^{\delta}_{i} - k^{\delta}_{i}).
\end{equation}
As vertices in $\preDelta{V}$ have at most $2^{\delta +1}$ edges
  and
  each edge of $\Ghat$ only appears in at most $\ceiling{\log 2 Q}$ sets $E^{i}$,
\begin{equation}\label{eqn:sparsify2b}
 \sum_{i,j \in S^{\delta}_{i}} \sizeof{\bdry{E^{i}}{\preDelta{C}^{i}_{j}}}
 \leq
 \ceiling{\log 2 Q}  2^{\delta + 1}  \sizeof{\preDelta{V}}.
\end{equation}
Combining \eqref{eqn:sparsify2a} with \eqref{eqn:sparsify2b}
  and \eqref{eqn:sumki}, we get
\[
 \sum_{i} t^{\delta}_{i}
\leq
  2 \ceiling{\log 2 Q} \sizeof{\preDelta{V}} + 2 l n,
\]
and so
\[
 \sum_{\delta , i} t^{\delta}_{i}
\leq
  2 \ceiling{\log 2 Q} n + 2 l n \ceiling{\log 2 n}
 \leq c_{8} n l\log n,
\]
for some constant $c_{8}$.
By now applying the analysis from the proof of Theorem~\ref{thm:sparsify},
  we may prove that $(Y.2)$ and $(Y.3)$ hold with probability
  at least $1-p$.
Of course, property $(Y.1)$ always holds.

To prove property $(Y.4)$,
  we note that the blow-up of a vertex $v$ is the sum
  of $1/d_{v}$ times the the blow-up of each of its edges.
We prove in Lemma~\ref{lem:expectedBlowUp} that the expectation of this
  sum is $1$, and in Lemma~\ref{lem:maxBlowUp} that each
  term is bounded by
\[
  \beta = \frac{1}{48 \log (3 n / p)^{2}}.
\]
If the variables were independent, we could apply Theorem~\ref{thm:chernoff}
  to prove it is unlikely that
  $v$ has blow-up greater than $2$.

However, the variables are not independent.
The blow-up of edges output by \boundedsparsify \ are independent.
But, the choice of a random pullback at line 4e introduces correlations
  in the blow-up of edges.
Fortunately, the blow-up of edges attached to $v$ have a negative association
  (as may be proved by Proposition~8 and Lemma~9 of
  Dubhashi and Ranjan~\cite{DubhashiRanjan}).
Thus, by Proposition~7 of~\cite{DubhashiRanjan}, we may still apply
  Theorem~\ref{thm:chernoff}, with $\epsilon = 1$ and $\mu = 1$ to show that the
\[
\prob{}{\blowup{\Gtil}{v} > 2}
\leq
e^{- 48 \log (3 n / p)^{2} / 3}.
\]
Applying a union bound over the vertices $v$, we see that $(Y.4)$
  hold with probability at least $1 - p/3$.

The analysis of the running time of \sparsifytwo \ is similar to
  the analysis of \sparsify, except for the work required to sub-divide
  sets in step 4b, which we now analyze.
Each time a vertex is removed from a set $\preDelta{C}^{i}_{j}$
  during the subdivision, the work required
  by a reasonable implementation is proportional to the degree
  of that vertex in graph $G^{i}$.
So, the work required to perform all the subdivisions over the course
  of the algorithm is at most
\[
  O \left(\sum_{\delta , i} 2^{\delta +1} \sizeof{S^{\delta}_{i}} \right).
\]
As
\[
 \bdry{E_{i}}{\preDelta{C}^{i}_{j}} \geq 2^{\delta }
\]
whenever we subdivide $\preDelta{C}^{i}_{j}$, we have
\[
  \sum_{j \in S^{\delta}_{i}}  \bdry{E_{i}}{\preDelta{C}^{i}_{j}}
 \geq 2^{\delta } \sizeof{S^{\delta}_{i}}.
\]
Now,  by \eqref{eqn:sparsify2b}
\[
\sum_{i}  2^{\delta } \sizeof{S^{\delta}_{i}}
\leq
 \ceiling{\log 2 Q}  2^{\delta + 1}  \sizeof{\preDelta{V}}
\leq
2  \ceiling{\log 2 Q}  \vol{\preDelta{V}}.
\]
Thus,
\[
\sum_{\delta, i} 2^{\delta +1} \sizeof{S^{\delta}_{i}}
\leq
4  \ceiling{\log 2 Q}  \vol{\preDelta{V}}
=
O (m \log (1/\epsilon)).
\]
The stated bound on the expected running time of \sparsifytwo \ follows.
\end{proof}

\begin{lemma}\label{lem:expectedBlowUp}
Let $\Gtil = (V, \Etil , \wtil )$
  be the graph output by \sparsifytwo \ on input $G = (V,E,w)$.
Then, for every $e \in E$,
\end{lemma}
\vspace{-0.2in}
\begin{equation}\label{eqn:blowup}
  \expec{}{{\rm \blowup{\Gtil }{e}}} \leq 1.
\end{equation}

\begin{proof}
We first observe that
\begin{equation}\label{eqn:eqblowup}
  \expec{}{\blowup{\Gtil }{e}} = 1.
\end{equation}
 holds for the graph $\Gtil$ output
  by \sample \, as it takes a weight-1 graph as input, selects a
  probability $p_{e}$ for each edge, and includes it at weight $1/p_{e}$
  with probability $p_{e}$.
As \unwtedsparsify  \ merely partitions its input into edge-disjoint
  subgraphs and then applies \sample \ to some of them, \eqref{eqn:eqblowup}
  holds for the output of \unwtedsparsify \ as well.

To show that \eqref{eqn:eqblowup} holds for the graph output by
  \boundedsparsify \, for each edge $e \in E$ and for each $i$ set
\[
w_{e}^{i}
=
\begin{cases}
1 & \text{if $e \in G^{i}$}
\\
0 & \text{otherwise}.
\end{cases}
\]
We have
\[
  w_{e} = \sum_{i} 2^{i} w_{e}^{i}.
\]
For the graph $\Gtil_{i}$ returned on line 2 of \boundedsparsify, let
  $\Gtil^{i} = (V, \Etil^{i}, \wtil^{i})$.
We have established that
\[
  \expec{}{\wtil^{i}_{e}} = w^{i}_{e}.
\]
So,
\[
 \expec{}{\blowup{\Gtil }{e}}
=
\expec{}{\frac{\sum_{i} 2^{i} \wtil^{i}_{e}}{w_{e}}}
=
 \frac{\sum_{i} 2^{i} \expec{}{\wtil^{i}_{e}}}{w_{e}}
=
 \frac{\sum_{i} 2^{i} w^{i}_{e}}{w_{e}}
=
1,
\]
establishing \eqref{eqn:eqblowup} for the output of \boundedsparsify .

Applying similar reasoning, we may establish \eqref{eqn:blowup} for
  the output of \sparsifytwo \ by proving that for each edge $e$ in
  each weight-1 graph $G^{i}$,
  the expected blow-up of $e$ in $\Gtil^{i}$ is at most $1$.
If $e$ is not on the boundary of a set $\preDelta{C}^{i}_{j}$,
  then $e$ will not appear in $\Gtil^{i}$ and so its blow-up will be zero.
If $e = (u,v)$ is on the boundary, then let $w_{e}$ denote the number
  of edges $e' = (u', v')$ for which $\pi (u) = \pi (u')$ and $\pi (v) = \pi (v')$.
If we let $H = (Y, F, y)$ and $\Htil = (Y, \Ftil , \ytil )$,
  then $w_{e} = y_{(\pi (u), \pi (v))}$.

Now, let $f$ be the edge $(\pi (u), \pi (v))$ in $H$.
We know that $\expec{}{\blowup{\Htil}{f}} = 1$.
If $f$ appears in $\Htil$, then the probability that edge $e$ is chosen
  in the random pullback is $1/w_{e}$.
As $f$ has weight $w_{e}$, we find
\[
  \expec{}{\blowup{\Gtil^{i}}{e}}
=
\frac{1}{w_{e}} \left(w_{e}   \expec{}{\blowup{\Htil^{i}}{f}}  \right)
= 1.
\]
\end{proof}

\begin{lemma}\label{lem:maxBlowUp}
Let $\Gtil = (V, \Etil , \wtil )$
  be the graph output by \sparsifytwo \ on input $G = (V,E,w)$.
Then, for every $(u,v) \in E$,
\end{lemma}
\vspace{-0.2in}
\begin{equation}\label{eqn:maxblowup}
  \blowup{\Gtil }{u,v} \leq
  \frac{\min (d_{u}, d_{v})}{48 \log (3 n /p)^{2}}.
\end{equation}

\begin{proof}
As in the proof of the previous lemma, we work our way though the algorithms
  one-by-one.
The graph produced by the  algorithm \sample \
  has blow-up at most $\min (d_{u}, d_{v}) / (16 \log (3/p))^{2}$
  for every edge $(u,v)$.
As \unwtedsparsify \ only calls \sample \ on subgraphs of its input graph,
  a similar guaranteed holds for the output of \unwtedsparsify.
In fact, as \unwtedsparsify \ calls \sample \ with $\phat < p/n$,
  every edge output by \unwtedsparsify \ actually has blow-up less than
\[
\min (d_{u}, d_{v}) / (16 \log (3 n /p))^{2}.
\]
As \boundedsparsify \ merely calls \unwtedsparsify \ on a collection of
  graphs that sum to $G$, the same bound holds on the blow-up of the
  graph output by \boundedsparsify  .

To bound the blow-up of edges in the graph output by \sparsifytwo  ,
  note that for every $i$ and every vertex $a$ in a graph $H^{i}$,
  the vertices $v$ of the original graph that map to $H^{i}$ under
  $\pi$ satisfy
\[
  d_{v} \geq 4 d_{a},
\]
where $d_{v}$ refers to the degree of vertex $v$ in the original graph
  and $d_{a}$ is the degree of vertex $a$ in graph $H^{i}$.
So, the blow-up of every edge $(u,v) \in E^{i}$ satisfies
\[
\blowup{\Gtil^{i}}{u,v} \leq
\frac{4 \min (d_{u}, d_{v})}{ (16 \log (3 n /p))^{2}}
=
\frac{\min (d_{u}, d_{v})}{ 48 \log (3 n /p)^{2}}
\]

We now measure the blow-up of edges relative to $\Ghat$ instead of
  $G$, which can only over-estimate their blow-up.
The lemma then follows from
\[
  \blowup{\Gtil}{u,v}
=
  \sum_{i} \frac{2^{-i} \blowup{\Gtil^{i}}{u,v}}{z_{u,v}}
\leq
\frac{\min (d_{u}, d_{v})}{ 48 \log (3 n /p)^{2}}
  \sum_{i} \frac{2^{-i}}{z_{u,v}}
=
\frac{\min (d_{u}, d_{v})}{ 48 \log (3 n /p)^{2}}.
\]

\end{proof}

\section{Final Remarks}\label{sec:conclusion}

Since the initial announcement \cite{SpielmanTengPrecon} 
  of our results, significant improvements 
  have been made in spectral sparsification. 
Spielman and Srivastava~\cite{SpielmanSrivastava} have
  proved that spectral sparsifiers with $O (n \log n / \epsilon^{2})$ edges
  exist, and may be found in time $\softO{m \log (n W / \epsilon)}$ where
  $W$ is the ratio of the largest weight to the smallest 
  weight of an edge in the input graph.
Their nearly-linear time algorithm relies upon the solution of a logarithmic
  number of linear systems in diagonally-dominant matrices.
Until recently, the only nearly-linear time algorithm for solving such systems
  was the algorithm in~\cite{SpielmanTengLinsolve}, which relied upon the constructions
  in this paper.
Recently, Koutis, Miller and Peng~\cite{KMP} have developed a faster
  algorithm that does not rely on the sparsifier construction of the present paper.
Their algorithm finds $\alpha $-approximate solutions to Laplacian linear systems
  in time $O (m \log^{2} n \log \alpha^{-1})$.
One may remove the dependence on $W$ in the running time of the
  algorithm of~\cite{SpielmanSrivastava} through the procedure
  described in Section~\ref{sec:weighted} of this paper.
Batson, Spielman and Srivastava~\cite{BatsonSpielmanSrivastava}
  have shown that sparsifiers with $O (n / \epsilon^{2})$ edges exist, and
  present a polynomial-time algorithm that finds these sparsifiers.
It is our hope that sparsifiers with so few edges may also be found
  in nearly-linear time.

Andersen, Chung and Lang~\cite{AndersenChungLang} 
  and Andersen and Peres~\cite{AndersenPeres} have
  improved upon some of the core algorithms we presented in~\cite{SpielmanTengCuts}
  and in particular have improved upon the algorithm \partition \ upon which
  we based \approxcut.
The algorithm of Andersen and Peres~\cite{AndersenPeres} is both significantly
  faster and saves a factor of $\log^{2} m$ in the conductance of the set it
  outputs.
In particular, it satisfies guarantee $(P.3)$ with the term
  $O (\tau^{2} / \log n)$ in place of our function $f_{1} (\tau )$.

\bibliographystyle{alpha}
\bibliography{precon}

\newcommand{\etalchar}[1]{$^{#1}$}
\begin{thebibliography}{BGH{\etalchar{+}}06}

\bibitem[ACL06]{AndersenChungLang}
Reid Andersen, Fan Chung, and Kevin Lang.
\newblock Local graph partitioning using pagerank vectors.
\newblock {\em Proceedings of the 47th Annual Symposium on Foundations of
  Computer Science}, pages 475--486, 2006.

\bibitem[AM01]{AchlioptasMcSherry}
Dimitris Achlioptas and Frank McSherry.
\newblock Fast computation of low rank matrix approximations.
\newblock In {\em Proceedings of the 33rd Annual {ACM} Symposium on Theory of
  Computing}, pages 611--618, 2001.

\bibitem[AP09]{AndersenPeres}
Reid Andersen and Yuval Peres.
\newblock Finding sparse cuts locally using evolving sets.
\newblock In {\em STOC '09: Proceedings of the 41st annual ACM symposium on
  Theory of computing}, pages 235--244, New York, NY, USA, 2009. ACM.

\bibitem[Axe85]{Axelsson}
O.~Axelsson.
\newblock A survey of preconditioned iterative methods for linear systems of
  algebraic equations.
\newblock {\em BIT Numerical Mathematics}, 25(1):165--187, March 1985.

\bibitem[BGH{\etalchar{+}}06]{SupportGraph}
M.~Bern, J.~Gilbert, B.~Hendrickson, N.~Nguyen, and S.~Toledo.
\newblock Support-graph preconditioners.
\newblock {\em SIAM J. Matrix Anal. \& Appl}, 27(4):930--951, 2006.

\bibitem[BH03]{SupportTheory}
Erik~G. Boman and Bruce Hendrickson.
\newblock Support theory for preconditioning.
\newblock {\em SIAM Journal on Matrix Analysis and Applications},
  25(3):694--717, 2003.

\bibitem[BK96]{BenczurKarger}
Andr{\'a}s~A. Bencz{\'u}r and David~R. Karger.
\newblock Approximating s-t minimum cuts in {O}(n{$^{2}$}) time.
\newblock In {\em Proceedings of The Twenty-Eighth Annual {ACM} Symposium On
  The Theory Of Computing ({STOC} '96)}, pages 47--55, May 1996.

\bibitem[Bol98]{BollobasMGT}
B{\'e}la Bollob{\'a}s.
\newblock {\em Modern graph theory}.
\newblock Springer-Verlag, New York, 1998.

\bibitem[BSS09]{BatsonSpielmanSrivastava}
Joshua~D. Batson, Daniel~A. Spielman, and Nikhil Srivastava.
\newblock Twice-{Ramanujan} sparsifiers.
\newblock In {\em Proceedings of the 41st Annual ACM Symposium on Theory of
  computing}, pages 255--262, 2009.

\bibitem[Che70]{Cheeger}
J.~Cheeger.
\newblock A lower bound for smallest eigenvalue of laplacian.
\newblock In {\em Problems in Analysis}, pages 195--199, In R.C. Gunning
  editor,, Princeton University Press, 1970.

\bibitem[Che89]{PaulChew}
Paul Chew.
\newblock There are planar graphs almost as good as the complete graph.
\newblock {\em J. Comput.~Syst.~Sci.}, 39:205--219, 1989.

\bibitem[Chu97]{Chung}
Fan R.~K. Chung.
\newblock {\em Spectral Graph Theory}.
\newblock CBMS Regional Conference Series in Mathematics. American Mathematical
  Society, 1997.

\bibitem[DR98]{DubhashiRanjan}
Devdatt Dubhashi and Desh Ranjan.
\newblock Balls and bins: a study in negative dependence.
\newblock {\em Random Structures and Algorithms}, 13(2):99--124, 1998.

\bibitem[DS91]{DiaconisStrook}
Persi Diaconis and Daniel Stroock.
\newblock Geometric bounds for eigenvalues of markov chains.
\newblock {\em The Annals of Applied Probability}, 1(1):36--61, 1991.

\bibitem[FK81]{FurediKomlos}
Z.~F{\"u}redi and J.~Koml{\'o}s.
\newblock The eigenvalues of random symmetric matrices.
\newblock {\em Combinatorica}, 1(3):233--241, 1981.

\bibitem[GR01]{GodsilRoyle}
Chris Godsil and Gordon Royle.
\newblock {\em Algebraic Graph Theory}.
\newblock Graduate Texts in Mathematics. Springer, 2001.

\bibitem[KMP10]{KMP}
Ioannis Koutis, Gary~L. Miller, and Richard Peng.
\newblock Approaching optimality for solving sdd systems.
\newblock March 2010.
\newblock Available at \texttt{http://arxiv.org/abs/1003.2958v1}.

\bibitem[KVV04]{KannanVempalaVetta}
Ravi Kannan, Santosh Vempala, and Adrian Vetta.
\newblock On clusterings: Good, bad and spectral.
\newblock {\em J. ACM}, 51(3):497--515, 2004.

\bibitem[LPS88]{LPS}
A.~Lubotzky, R.~Phillips, and P.~Sarnak.
\newblock {Ramanujan} graphs.
\newblock {\em Combinatorica}, 8(3):261--277, 1988.

\bibitem[Mar88]{Margulis}
G.~A. Margulis.
\newblock Explicit group theoretical constructions of combinatorial schemes and
  their application to the design of expanders and concentrators.
\newblock {\em Problems of Information Transmission}, 24(1):39--46, July 1988.

\bibitem[Moh91]{Mohar91Laplacian}
Bojan Mohar.
\newblock The {L}aplacian spectrum of graphs.
\newblock In {\em Graph Theory, Combinatorics, and Applications}, pages
  871--898. Wiley, 1991.

\bibitem[Rag88]{Raghavan}
Prabhakar Raghavan.
\newblock Probabilistic construction of deterministic algorithms: Approximating
  packing integer programs.
\newblock {\em J. Comput.~Syst.~Sci.}, 37:130--143, 1988.

\bibitem[SJ89]{JerrumSinclair}
Alistair Sinclair and Mark Jerrum.
\newblock Approximate counting, uniform generation and rapidly mixing {Markov}
  chains.
\newblock {\em Information and Computation}, 82(1):93--133, July 1989.

\bibitem[SS08]{SpielmanSrivastava}
Daniel~A. Spielman and Nikhil Srivastava.
\newblock Graph sparsification by effective resistances.
\newblock In {\em Proceedings of the 40th annual ACM Symposium on Theory of
  Computing}, pages 563--568, 2008.

\bibitem[ST04]{SpielmanTengPrecon}
Daniel~A. Spielman and Shang-Hua Teng.
\newblock Nearly-linear time algorithms for graph partitioning, graph
  sparsification, and solving linear systems.
\newblock In {\em Proceedings of the thirty-sixth annual {ACM} Symposium on
  Theory of Computing}, pages 81--90, 2004.
\newblock Full version available at
  \texttt{http://arxiv.org/abs/cs.DS/0310051}.

\bibitem[ST08a]{SpielmanTengCuts}
Daniel~A. Spielman and Shang-Hua Teng.
\newblock A local clustering algorithm for massive graphs and its application
  to nearly-linear time graph partitioning.
\newblock {\em CoRR}, abs/0809.3232, 2008.
\newblock Available at \texttt{http://arxiv.org/abs/0809.3232}. Submitted to
  SICOMP.

\bibitem[ST08b]{SpielmanTengLinsolve}
Daniel~A. Spielman and Shang-Hua Teng.
\newblock Nearly-linear time algorithms for preconditioning and solving
  symmetric, diagonally dominant linear systems.
\newblock {\em CoRR}, abs/cs/0607105, 2008.
\newblock Available at \texttt{http://www.arxiv.org/abs/cs.NA/0607105}.
  Submitted to SIMAX.

\bibitem[Tar75]{Tarjan}
R.~E. Tarjan.
\newblock Efficiency of a good but not linear set union algorithm.
\newblock {\em Journal of the ACM}, 22(2):448--501, 1975.

\bibitem[TB97]{TrefethenBau}
L.~N. Trefethen and D.~Bau.
\newblock {\em Numerical Linear Algebra}.
\newblock SIAM, Philadelphia, PA, 1997.

\bibitem[Tre05]{Trevisan}
Lucan Trevisan.
\newblock Approximation algorithms for unique games.
\newblock {\em Proceedings of the 46th Annual IEEE Symposium on Foundations of
  Computer Science}, pages 197--205, Oct. 2005.

\bibitem[Vu07]{Vu}
Van Vu.
\newblock Spectral norm of random matrices.
\newblock {\em Combinatorica}, 27(6):721--736, 2007.

\end{thebibliography}

\end{document}